\documentclass[aip,reprint,jcp,citeautoscript,amsmath]{revtex4-1}

\usepackage{float}

\usepackage{ textcomp }
\usepackage{amsmath} 
\usepackage{siunitx}
\usepackage[version=4]{mhchem}
\usepackage{wrapfig}
\usepackage{hyperref}
\usepackage{amsfonts}
\usepackage{amssymb}
\usepackage{mathtools}
\usepackage{subcaption}
\usepackage{float}
\usepackage[T1]{fontenc}
\DeclareSIUnit{\molar}{M}

\draft 

\begin{document}

\title{Many-body attractions do not stabilize gas-liquid phase separation in aqueous dispersions of charged colloids within the Poisson-Boltzmann framework}



\author{Thijs ter Rele}
\email[]{t.r.terrele@uu.nl}
\affiliation{Soft Condensed Matter \& Biophysics, Debye Institute for Nanomaterials Science, Utrecht University, Princetonplein 1,3584 CC Utrecht, the Netherlands}

\author{Ren\'{e} van Roij}
\email[]{r.vanroij@uu.nl}
\affiliation{Institute for Theoretical Physics, Utrecht University, Princetonplein 5, 3584 CC Utrecht, the Netherlands}

\author{Marjolein Dijkstra}
\email[]{m.dijkstra@uu.nl}
\affiliation{Soft Condensed Matter \& Biophysics, Debye Institute for Nanomaterials Science, Utrecht University, Princetonplein 1,3584 CC Utrecht, the Netherlands}


\date{\today}

\begin{abstract}
Attractive three-body interactions have been reported for like-charged colloids in low-salt suspensions, based on both finite-element Poisson-Boltzmann calculations and direct experimental measurements, and have been proposed as a mechanism to drive colloidal clustering. However, these Poisson-Boltzmann calculations typically neglect charge regulation and higher-order many-body effects. Here, we construct machine-learned many-body interaction potentials for charge-regulating colloids, trained on finite-element Poisson-Boltzmann calculations, to accurately capture three-body and higher-order contributions. We find that the three-body contribution to the many-body potential as obtained from Poisson–Boltzmann calculations on isolated colloid triplets is strongly attractive, consistent with previous work, whereas the four-body contribution for an equilateral pyramid configuration of four colloids is repulsive. We then construct machine-learned many-body potentials for charged colloids using finite-element Poisson-Boltzmann calculations on clusters of 13 colloids, and find that the incorporation of higher-body interactions weakens the cohesive nature of the interactions. We identify a parameter regime exhibiting gas-liquid or gas-solid phase separation using the machine-learned potentials in molecular dynamics simulations. However, when we include clusters of 48 colloids in the training data, the cohesion diminishes further, and molecular dynamics simulations using these potentials no longer include broad phase separation in aqueous dispersions of charged colloids. Finally, we compute the potential of mean force of pairs, triplets  and quartets of colloids using primitive model simulations. We find that the resulting potentials are in good agreement with those obtained from the Poisson-Boltzmann calculations, thereby supporting the validity of the Poisson-Boltzmann approach for determining many-body interactions. 
\end{abstract}


\maketitle 

\section{Introduction}

The interactions between Brownian charged colloids suspended in a liquid  electrolyte are commonly  described using a combination of the Poisson equation and the Boltzmann distribution. The Poisson equation relates the charge density, stemming from both the suspended colloids and the dissolved  microscopic ions in a dielectric continuum, to the resulting electrostatic potential, while the Boltzmann distribution describes the thermal equilibrium ion concentrations  in that potential at the mean-field level. Together, they form the Poisson-Boltzmann (PB) equation, which takes the form of a nonlinear (partial) differential equation for the electrostatic potential in a given configuration of charged colloids, with the colloidal (surface) charges acting as the source term. Only a few analytic solutions exist, perhaps most notably the Gouy-Chapman solution for a single homogeneously charged plane in contact with a half space of electrolyte.\cite{Gouy1910, Chapman1913} In the regime  of low colloidal charges, and hence small electrostatic potentials, the Poisson-Boltzmann equation can be linearized, such that (approximate) solutions have been constructed analytically in a wider range of  geometries. For instance, in the case of \emph{two} suspended like-charged spheres at center-to-center distance $r$, this linearisation procedure was successfully used to derive a repulsive screened-Coulomb potential $\propto \exp(-\kappa r)/r$, where $\kappa^{-1}$  is the Debye screening length of the electrolyte. This result lies at the heart of the Derjaguin-Landau-Verwey-Overbeek (DLVO) potential \cite{Derjaguin1941, Verwey1948book}, which includes apart from  this electrostatic Yukawa-like pair potential and a hard-sphere repulsion also an attractive Van der Waals interaction. 
The latter can often be neglected  when index-matching between the colloids and solvent is applied, leaving the electrostatic and hard-sphere repulsion as the dominant interactions. The DLVO potential has long served as a cornerstone of colloid science, at least for aqueous electrolytes and low-valency ionic species.

Nevertheless, in the 1990's, several experiments raised doubts about the validity of the DLVO potential at extremely low salt concentrations, where the  screening length $\kappa^{-1}$ becomes of the order of the colloid radius. Under these conditions, indications of effective attractions between like-charged particles were reported,\cite{Tata1992, Ito1994, Kepler1994, Larsen1997, Wang2024} although some of these findings  were later questioned due to possible experimental artifacts\cite{Palberg1994_tata, Squires2000_brenner}  and artifacts related to optical distortions influencing the pair potential measurements.\cite{baumgartl-2006} 
In spite of this controversy, these observations stimulated extensive theoretical work on like-charge attractions, particularly focusing on whether such effects could give rise to ``broad'' phase coexistence between dense and dilute phases in suspensions of charged colloidal spheres. 

Some Poisson-Boltzmann  calculations have attributed  the broad phase coexistence to ``volume terms'' \cite{vanRoij1997, vanRoij1999,Warren2000, Denton2000, Zoetekouw2006prl,Zoetekouw2006}, which  arise as a cohesive many-body free-energy contribution to the density-dependent self-energy of colloids with their own double layer.  This contribution stems from linearisation of the Poisson-Boltzmann equation about the density-dependent Donnan potential, rather than about zero potential, such that the effective thickness of the screening layer depends on the colloid density. 
For systems with long screening lengths $\kappa^{-1}$, on the order of the colloid particle radius, this cohesion was shown to become sufficiently strong to drive phase separation. 
Alternatively, within nonlinear Poisson-Boltzmann theory, finite-element methods   provide a powerful tool for studying colloidal interactions. The Poisson-Boltzmann equations have been solved  using finite-element approaches in one or two dimensions since the 1980s,\cite{Alexander1984, James1985} with later extensions  to three dimensions,\cite{Holst2000, Holst2012} enabling detailed investigations of charge renormalization and dielectric effects in colloidal suspensions.\cite{Schlaich2023}

Using finite-element nonlinear Poisson-Boltzmann calculations, an attractive three-body interaction  was identified that surprisingly  depends primarily on the total perimeter of  the triangle formed by centers of the three colloids.\cite{Russ2002} Shortly thereafter, direct experimental measurements confirmed the presence of a similar attractive three-body contribution.\cite{Dobnikar2004, Brunner2004, Merrill2009} 
In addition,  Monte Carlo  simulations revealed an attractive many-body interaction  between charged nanoparticles, which becomes stronger at lower salt 
concentrations.\cite{Zhang2016} 
This many-body attraction was predicted to significantly influence the phase behavior of charged colloids,\cite{Dobnikar2003} and  free-energy calculations demonstrated that it can induce extended regions of phase coexistence despite purely repulsive pair interactions.\cite{Hynninen2004} However, these studies were all based on effective two- and three-body potentials that neglected higher-order many-body contributions. Interestingly, simulations of asymmetric primitive models, which do not rely on the input of effective interactions because of an explicit representation of all microscopic ions,  indicate that such higher-order terms are predominately repulsive and tend to cancel out the attractive three-body interactions. This is consistent with  the absence of observations of broad  phase coexistence regions  in primitive model simulations.\cite{Hynninen2005}

Russ {\em et al}. captured interaction terms up to four-body contributions using finite-element Poisson-Boltzmann calculations.\cite{Russ2002} Extending this approach to higher-order interactions  becomes increasingly more difficult, as the configuration space grows rapidly. In recent previous work, we used a Machine Learning (ML) approach to construct effective many-body interaction potentials for charged colloids.\cite{terRele2025, terRele2026} In this treatment, forces obtained from many-colloid configurations in primitive model simulations are fitted to an ML potential expressed in terms of two- and three-body Behler-Parrinello symmetry functions.\cite{Behler2007} Although the functional form explicitly  contains only   up to three-body terms, the training procedure allows the ML model to implicitly encode higher-order many-body interactions. This is because the total many-body potential is constructed as a sum of local particle  contributions, where the contribution of  each particle depends on the local arrangement of  neighboring particles within a finite cutoff radius. As a result, the ML model can effectively  capture collective many-body effects beyond explicit three-body interactions, including an approximate dependence of the interactions on the local particle density. 

In this work, we generate many-body interaction potentials using finite-element Poisson-Boltzmann calculations on isolated pairs, triplets, and quadruplets of colloids as described by Russ {\em et al}.,\cite{Russ2002} and  compare them with both the DLVO potential and machine-learned potentials generated using our ML framework based on Poisson-Boltzmann calculations of up to 48 colloids.\cite{terRele2025} We then employ these potentials in simulations to assess whether such colloidal systems exhibit broad phase coexistence.
 
\section{Method}

\begin{figure}[h!]
\centering
\includegraphics[width=\linewidth]{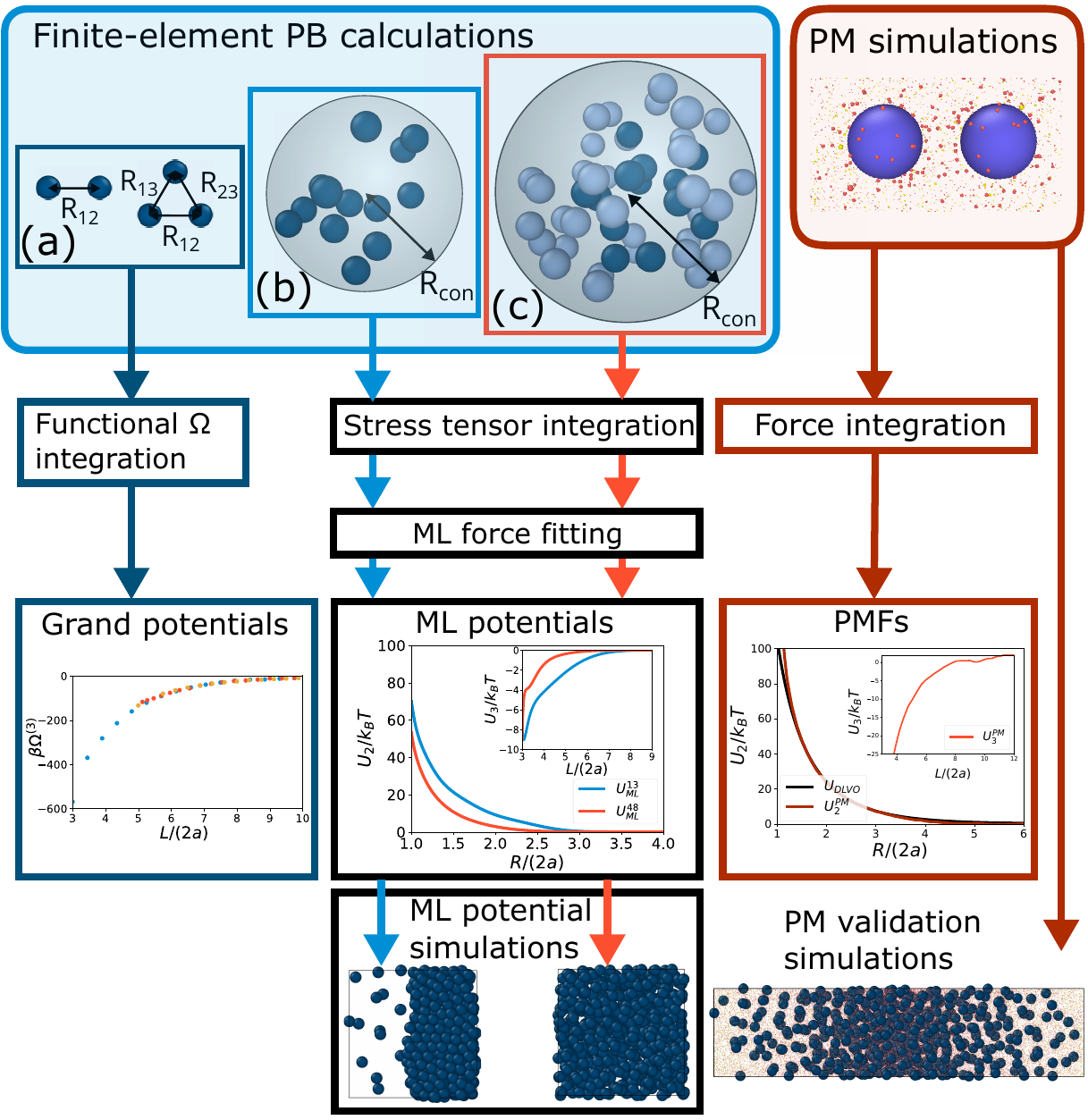}
\caption{\textbf{Overview of the workflow used to generate the potentials, figures and simulations presented in this work.} Through PB calculations of systems containing different  numbers of colloids, we obtain many-body interaction potentials. (a) Three colloids arranged in an equilateral configuration, used for the direct calculation of $\Omega^{(3)}$. (b) Thirteen colloids confined within a sphere of radius $R_{con}$, with all colloids included in the training of the ML potential $U_{ML}$. (c) Forty-eight colloids confined within a sphere of radius $R_{con}$, where only the  thirteen colloids  closest to the center of the confining sphere (shown in dark blue) are  used for training the ML potential $U_{ML}$.  } 
\label{fig:workflow}
\end{figure}

We employ several computational and simulation methods to investigate interactions between charged colloids in low-salt suspensions, with  particular emphasis on the many-body character of these interactions. In this section, we describe how the colloidal interaction potentials are obtained using these different approaches and clarify how the resulting potentials are related to one another. Figure \ref{fig:workflow} provides an overview of the methods used to obtain the results supporting the conclusions of this work.

\subsection{Poisson-Boltzmann theory}\label{sec:PB-theory}
In this work, we perform numerical finite-element calculations on fixed configurations of $N$ non-overlapping charged colloidal spheres of hard-core radius $a$ suspended in an aqueous 1:1 electrolyte solution at room temperature $T$. We  refer to the region filled by the electrolyte as $\mathbf{r} \in G$, i.e. the space not occupied by the colloids.  Its boundary, corresponding to the colloid surfaces $\partial G_i$, is denoted by the set $\partial G=\partial G_1\cup\partial G_2\cup\cdots\cup\partial G_N$, where $i \in 1,...,N$ labels the individual  colloids. The geometry of the system is thus fully determined by the center-of-mass coordinates $\{\mathbf{R}_i\}$ of the colloids.  
The electrostatic potential $(k_BT/e)\phi(\mathbf{r})$ in $G$, with $k_B$ the Boltzmann constant and $e$ the elementary charge, is described within the  Poisson-Boltzmann framework through the Boltzmann distribution $\rho_{\pm}(\mathbf{r}) = \rho_s \exp{\left[\mp \phi(\mathbf{r}) \right]}$,  where $\rho_s$ is the salt concentration in a bulk salt reservoir that is in osmotic equilibrium with the suspension. 
In the Poisson-Boltzmann framework, ions are treated as point particles with zero radius. 
The dimensionless electrostatic potential $\phi({\bf r})$ in $G$ is related to the local ionic charge density $\rho_c(\mathbf{r}) = \rho_{+}(\mathbf{r}) -\rho_{-}(\mathbf{r})$ via 
the Poisson equation $\nabla^2 \phi(\mathbf{r}) = - 4 \pi \lambda_B\rho_c(\mathbf{r})$, where the Bjerrum length $\lambda_B=e^2/4\pi\epsilon k_BT$  sets  the strength of the electrostatic interactions in the electrolyte. Here $\epsilon$ is the static dielectric constant of the solvent. Combining the Poisson equation with the ionic Boltzmann distributions yields the Poisson-Boltzmann equation
\begin{equation}
\nabla^2 \phi(\mathbf{r}) = \kappa^2 \sinh \phi(\mathbf{r}) , \ \mathbf{r} \in G, 
\label{eq:PB}
\end{equation}
where $\kappa = \sqrt{8 \pi \lambda_B \rho_s}$ is the inverse Debye screening length, a property of the salt reservoir.  

To solve the nonlinear Poisson-Boltzmann  equation  (\ref{eq:PB}), appropriate  boundary conditions are required.  
These conditions follow from  Gauss' law and the assumption that the electric potential vanishes in the reservoir, i.e. far away at infinite distances from the nearest colloid.  For a system of colloids of radius $a$ with a surface charge density $\sigma(\mathbf{r})$ at ${\bf r}\in\partial G$, the boundary conditions read
\begin{align}
\phi(\mathbf{r}) &= 0,& \ |\mathbf{r}| \rightarrow \infty; \nonumber \\ 
\mathbf{n}_i \cdot \nabla \phi(\mathbf{r}) &= -4 \pi \lambda_B \sigma(\mathbf{r}),& \  \mathbf{r} \in \partial G_i ,
\label{eq:PB-BC}
\end{align}
where $\mathbf{n}_i$ is the unit vector,  normal to the surface $\partial G_i$ of colloid $i$, and pointing into the region $G$. A common choice is to assume a constant surface charge density $e\sigma({\bf r}) = Ze/(4\pi a^2)$, with $Z$ the colloid  valency, as was done by Russ {\em et al}.\cite{Russ2002} In part of this work we will use this constant-charge boundary condition as well. However, a more realistic description of the colloid charge accounts for charge regulation, where the surface charge  depends on the local electrostatic potential. \cite{ninham-1971, biesheuvel-2004A} In this work, we therefore also use a model with the local surface charge depending on the equilibrium of the association-dissociation reaction $\mathrm{AH} \leftrightharpoons \mathrm{A^-} + \mathrm{H^+}$, 
where $\mathrm{A^-}$ represents a  charged surface group on the colloid and $\mathrm{H^+}$ the released proton. We implement this using a Langmuir adsorption isotherm 
\begin{equation}
    f(\mathbf{r}) = \frac{1}{1 + 10^{
  \mathrm{p}K -\mathrm{pH}}\exp{\left( -\phi(\mathbf{r}) \right)}}, \,\,\,\,\,{\bf r}\in\partial G,
    \label{eq:alpha_Bies}
\end{equation}
with $f(\mathbf{r})$ the degree of ionization of the chargeable sites on the surface,  $\mathrm{pH}$ the global acidity of the suspension, and $\mathrm{p}K$  the negative logarithm of the reaction constant $K = 10^{-\mathrm{p}K}\text{M}$. 
Thus, we assume here that the colloid surfaces only consist of acid groups and that they carry a negative charge.
From the ionization degree, we retrieve the colloid surface charge density  e$\sigma(\mathbf{r}) = -e\Gamma f(\mathbf{r})$ for ${\bf r}\in\partial G$, with $\Gamma = M/(4\pi a^2)$ the surface density of chargeable sites, where $M$ is the total number of sites on a single colloid.\cite{Everts2016,Kubincov2020}
With these boundary conditions and for given parameters $M$ and $K$ that characterize the surface chemistry,   Eqs. (\ref{eq:PB}) with (\ref{eq:PB-BC}) can be solved numerically for arbitrary colloid configurations  using  finite-element methods. In this work, we employ the  \textsc{COMSOL} Multiphysics software package for this purpose,\cite{comsol} for either the fixed-charge or for the charge-regulating model. 
Using the \textsc{COMSOL} Multiphysics software package, we solve Eq. (\ref{eq:PB}) on a finite-element mesh, while self-consistently enforcing the standard charge-regulation boundary conditions given by Eqs. (\ref{eq:PB-BC}) and (\ref{eq:alpha_Bies}) through an iterative Newton–Raphson scheme.

\subsection{Grand potential}\label{sec:grandpot}
We wish to compare the finite-element Poisson-Boltzmann calculations to the DLVO potential. To this end, we determine the grand potential $\Omega_N(\{ {\bf R}_i\})$ of an $N$-body colloid  configuration from the electrostatic potential $\phi(\mathbf{r})$, following a procedure similar to that of Russ {\em et al}.\cite{Russ2002} However, in our charge-regulation case the colloidal surface charge is no longer constant, which introduces  an additional grand-potential contribution associated with the entropy and the enthalpy of the association-dissociation equilibrium of the chargeable groups. This effect was for instance accounted for by  Everts {\em et al}, \cite{Everts2016} and leads for charge-regulating colloids to the grand potential functional $\Omega[\rho_+,\rho_-,f]$ given in units of $k_BT=\beta^{-1}$ by
\begin{align}
\beta \Omega=& \sum_{\alpha = \pm} \int_{G} d^3 \mathbf{r} \rho_\alpha({\bf r})\left(\ln{(\rho_\alpha({\bf r}) /\rho_s)} - 1 \right)  \nonumber
\\ +& \int_{G} d^3 \mathbf{r}  \left[ \frac{1}{2} Q(\mathbf{r})\phi(\mathbf{r}) + 2 \rho_s \right]  \nonumber
\\ + &\,\Gamma\sum_i \int_{\partial G_i} d^2 \mathbf{r}  \bigg[ f(\mathbf{r}) \Big(\ln f(\mathbf{r})+\ln 10^{\mathrm{pK}-\mathrm{pH}}\Big)   \nonumber
\\ &\hspace{2cm}+ (1-f({\bf r})) \ln(1-f({\bf r}))\bigg],
\label{eq:grandpot-large}
\end{align}
where  $Q(\mathbf{r}) =\rho_c(\mathbf{r}) -\Gamma f(\mathbf{r})\sum_i\delta\left(|\mathbf{r} -  \mathbf{R}_i | -a\right)$ is the total (ion and surface) charge density. The different terms correspond, respectively, to the  ideal-gas free energy of the mobile ions, the total electrostatic (Coulomb) energy, an irrelevant constant shift $\propto 2\rho_s$ such that $\Omega=0$ for $N=0$, an entropic contribution $\propto f\ln f$ of the charged surface sites together with a free-energy contribution $\propto f\ln 10^{\mathrm{pK}}$ that accounts for the non-electrostatic bond-breaking due to deprotonation, and finally the entropy contribution $\propto (1-f)\ln(1-f)$ of neutral groups.  
Here we treated the surface charge (just like the cations and anions) grand-canonically, where mass action dictates that the difference of the chemical potential of the charged and uncharged surface groups equals $k_BT\ln 10^{-\mathrm{pH}}$, and chose an additive offset such that the grand potential of an uncharged surface group vanishes. 
One checks that the Euler-Lagrange equations $\delta\Omega/\delta\rho_\pm({\bf r})=0$ and $\delta\Omega/\delta f({\bf r})=0$ lead to Eqs.~(\ref{eq:PB}) and (\ref{eq:alpha_Bies}), respectively, for a given colloid configuration $\{{\bf R}_1,\cdots,{\bf R}_N\}$. 
Note that these are standard charge-regulation equations.\cite{Trefalt2015} 
After solving for $\phi({\bf r})$, and hence $\rho_\pm({\bf r)}$ and $f({\bf r})$, they can be inserted into the variational functional of Eq.(\ref{eq:grand-potential}) to obtain the equilibrium grand potential  for suspensions of charge-regulating colloids
\begin{align}
\beta \Omega_N &(\{\mathbf{R}_i \})= 
\rho_s\int_{G} d^3 \mathbf{r} \left[ \phi(\mathbf{r}) \sinh{\phi}(\mathbf{r}) -2\cosh{\phi}(\mathbf{r}) +2 \right] \nonumber
\\ &+ \Gamma \sum_i  \int_{\partial G_i} d^2  \mathbf{r} [\ln(1-f({\bf r})) + \frac{1}{2} \phi(\mathbf{r}) f (\mathbf{r})].
\label{eq:grand-potential}
\end{align}
For calculations on constant-charge colloids, the final line of Eq.(\ref{eq:grandpot-large}) is to be omitted, since it consists of the free energy contributions from the association-dissociation reaction and hence, there is no dependence on $f({\bf r})$. Still,  the Euler-Lagrange equations $\delta\Omega/\delta\rho_\pm({\bf r})=0$ lead for ${\bf r}\in G$ to the Boltzmann distributions $\rho_\pm({\bf r})=\rho_s\exp(\mp\phi({\bf r}))$ for the ions. For colloids with constant  surface charge density $e\sigma$, the equilibrium grand potential then reads
\begin{align}
\beta \Omega_N (\{\mathbf{R}_i \})&= 
\rho_s\int_{G} d^3 \mathbf{r} \left[ \phi(\mathbf{r}) \sinh{\phi}(\mathbf{r}) -2\cosh{\phi}(\mathbf{r}) +2 \right] \nonumber
\\ &+ \frac{\sigma}{2}  \sum_i  \int_{\partial G_i} d^2  \mathbf{r}   \phi(\mathbf{r}) .
\label{eq:grand-potential-constant-charge}
\end{align}
Thus, both for charge-regulating and constant-charge colloids, $\Omega_N(\{\mathbf{R}_i \})$ can be explicitly evaluated for any 
finite-element solution of the Poisson-Boltzmann equation by numerically integrating over all volume- and surface elements. We note that the two final terms of Eqs.~(\ref{eq:grand-potential}) and (\ref{eq:grand-potential-constant-charge}) have opposite signs: positive in the constant-charge case since $\sigma$ and $\phi({\bf r})$ have the same sign, and negative in the charge-regulation case since the fraction $-\sigma({\bf r})/\Gamma=f({\bf r})>0$ and $\phi({\bf r})<0$. The reason of this difference can be traced back to the (implicit) grand-canonical treatment of the surface charge in the latter case, which gives rise to a Legendre transformation of the relevant thermodynamic potential. For later convenience, this distinct treatment is not reflected by a different notation. 

In both cases, the equilibrium grand potential $\Omega_N(\{\mathbf{R}_i \})$ can next be decomposed into one-body, two-body, and higher-order many-body contributions by subsequently considering $N=1$, $N=2$, etc. For a translationally invariant system of $N$ colloids at positions $\{\mathbf{R}_i \}$  with $i = 1,..., N$, the  grand potential $\Omega_N (\{\mathbf{R}_i \})$ can be written  as
\begin{equation}
    \Omega_N(\{\mathbf{R}_i \}) = N \Omega_1 + \sum_{i<j} \Omega^{(2)}(ij)  + \hspace{-2mm}\sum_{i<j<k} \hspace{-2mm}\Omega^{(3)}(ijk)  + \cdots,
    \label{eq:om_terms}
\end{equation}
where $\cdots$ denotes four-body and higher-order terms.  For  $N=1$, the grand potential reduces to  the one-body contribution $\Omega_1$, corresponding to the self-energy of a single colloid in an infinitely large electrolyte reservoir. For $N = 2$, the two-body interaction potential $\Omega^{(2)}(R_{12}) \equiv \Omega^{(2)}(12)$  with colloid-colloid separation $R_{12} = |\mathbf{R}_1 -\mathbf{R}_2 |$, is defined from Eq. (\ref{eq:om_terms}) as
\begin{equation}
    \Omega^{(2)}(12) = \Omega_2 (12) - 2\Omega_1, 
\end{equation}
which vanishes in the limit  $R_{12} \rightarrow \infty$.
Similarly, the three-body contribution $\Omega^{(3)}(R_{12}, R_{13}, R_{23}) \equiv \Omega^{(3)}(123)$ is defined as
\begin{align}
        \Omega^{(3)}(123)  =  \Omega_3 (123) &\hspace{-1mm}- \hspace{-1mm}\Omega_2 (12)  \hspace{-1mm}- \hspace{-1mm}\Omega_2 (13) 
        \hspace{-1mm}-\hspace{-1mm} \Omega_2 (23) \hspace{-1mm}- \hspace{-1mm}3\Omega_1, 
\end{align}
and vanishes when any pairwise separation  $R_{ij} \rightarrow \infty$.

\subsection{Colloid forces}
To perform molecular dynamics simulations, one must determine the forces acting on the colloids  rather than the grand potential of the dispersion. This can be achieved by integrating the stress tensor $\mathbf{T}$ over the colloid surface, following the method of Von Gr\"{u}nberg and Mbamala,\cite{Grunberg2001} who write ${\bf T}$ as a sum of the (isotropic) osmotic stress tensor $2\rho_sk_BT(\cosh\phi({\bf r})-1)\mathbf{I}$ and the Maxwell stress tensor 
$(\epsilon/4\pi)(\frac{1}{2}E^2{\bf I}-{\bf E}{\bf E})$. Here $\mathbf{I}$ is the $3\times 3$ identity tensor and ${\bf E}=-(k_BT/e)\nabla\phi$ the electric field. 
We thus obtain  
\begin{align}
\beta {\mathbf{T}} &= \frac{\kappa^2}{4\pi \lambda_B} \left(\cosh\phi -1 \right) \mathbf{I} \nonumber
\\ & +\frac{1}{8\pi \lambda_B} \left((\nabla \phi)^2 \mathbf{I} -2 (\nabla \phi)(\nabla\phi) \right),
\end{align}
where we used  $\kappa^2 = 8 \pi \lambda_B \rho_s$ for later convenience. 
Once the Poisson-Boltzmann-Langmuir equations (\ref{eq:PB})-(\ref{eq:alpha_Bies}) have been solved for given configuration $\{{\bf R}_i\}$, or the Poisson-Boltzmann equation for the constant-charge case, the effective force ${\bf F}_i$ acting on colloid $i$ can be obtained by integrating the stress tensor over its surface, 
\begin{align}
\beta \mathbf{F}_i &=  \int_{\partial G_i} d^2 \mathbf{r} \beta {\mathbf{T}}\cdot  \mathbf{n}_i \nonumber
\\ &=\int_{\partial G_i} d^2 \mathbf{r} \bigg[ \frac{\kappa^2}{4\pi \lambda_B} \left(\cosh\phi -1 \right) \mathbf{I} \nonumber
\\&+\frac{1}{8\pi \lambda_B} \left((\nabla \phi)^2 \mathbf{I} -2 (\nabla \phi)(\nabla\phi) \right)\bigg] \cdot\mathbf{n}_i.
\label{eq:force-integral}
\end{align}
By using the colloid radius as unit of length, one checks that the resulting dimensionless force $\beta  {\bf F}_i a$ is only a function of the dimensionless system parameters $\kappa a$ (setting the screening strength) and $a/\lambda_B$ (setting the temperature), while $\phi({\bf r})$ also depends on the valency $Z$ for constant-charge particles and  on the dimensionless combination $\mathrm{pK}-\mathrm{pH}$ and the number of chargeable sites on an individual colloidal sphere $M\equiv 4\pi a^2\Gamma$ for charge-regulating particles.

\subsection{Machine learning}\label{sec:ML-procedure}
To determine whether phase separation occurs in systems of colloids, one needs to account for the complex interplay  of two-, three-, four-, and higher-body interactions. We capture these many-body effects by fitting the interaction forces obtained from  large groups of colloids to an ML model based on two-body and three-body Behler-Parrinello symmetry functions. In this way, higher-order interactions are incorporated into effective two-body and three-body potentials, which can be analyzed further and employed in large-scale simulations.

To construct a machine-learned potential for a given set of parameters, we first generate training data using  finite-element Poisson-Boltzmann calculations for systems of $N = 13$ colloids. 
In these calculations, the colloids are placed near the center of a large spherical cell of radius $R = 35a$, where we impose $\phi(R) = 0$ at the boundary, corresponding to the first boundary condition of Eq. (\ref{eq:PB-BC}). To sample a range of packing fractions, the colloids are confined within  a smaller sphere of  radius  $R_{con}$, which we varied  from $3.7 a$ to $12.0 a$, corresponding to packing fractions $\eta = 0.26$ down to $0.0075$.
In Fig. \ref{fig:workflow}(b) this setup is illustrated, with the 13 colloids, used for  training the potential, randomly placed at fixed positions within this sphere. 
To account for higher packing fractions, we additionally import configurations with packing fractions ranging from $\eta = 0.26$ up to $\eta = 0.65$, on which we  perform the  calculations. 
For each configuration, we set the gradient of the electric potential at the surface of each colloid $i$ to $\mathbf{n}_i \cdot \nabla \phi(\mathbf{r}) = -4 \pi \lambda_B\sigma(\mathbf{r}) $, implementing the second boundary condition of Eq.~(\ref{eq:PB-BC}). 
When charge-regulating colloids are used, an iterative procedure is applied to enforce Eq.~(\ref{eq:alpha_Bies}). 
For each parameter set, we perform finite-element Poisson-Boltzmann  calculations for 180 independent configurations of 13 colloids and compute the forces on each colloid using Eq. (\ref{eq:force-integral}). These forces, together with the corresponding colloid coordinates, form the training set for constructing the machine-learned interaction potentials, using a linear-regression-based force-matching procedure. The approach of constructing these ML potentials is described in detail in Ref. \citenum{terRele2025}. Once trained, these ML potentials enable efficient large-scale  simulations of the colloids-only system fully characterized by the center-of-mass coordinates $\{{\bf R}_i\}$.

\section{Results}

\subsection{Two-body interactions} \label{sec:twocol}

\begin{figure*}
\centering
\includegraphics[width=0.999\textwidth]{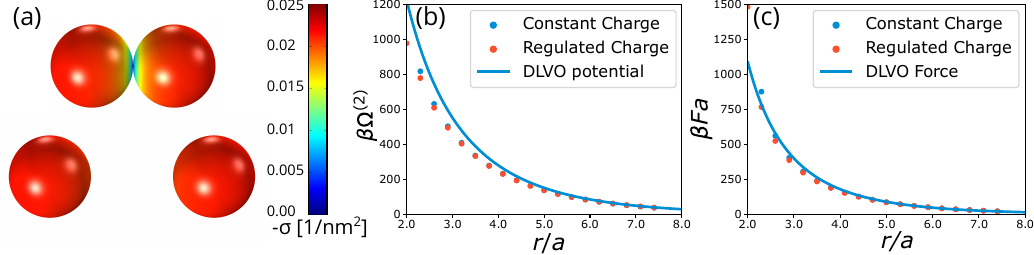}
\caption{\textbf{Interaction between two isolated colloids} at an effective temperature $a/\lambda_B =92.7 $, screening parameter $\kappa a  = 0.392$, and charge valency $Z = 1200$. Charge regulation is implemented using  $M = 10^5$, $\text{pH} = 7.0$, and $\text{p}K = 6.23$. (a) Local charge density $\sigma(\mathbf{r})$ on two charge-regulating colloids at center-to-center separations  $R = 2a$ and $R = 4a$. (b) Two-body grand potential $\beta \Omega^{(2)}$ for constant-charge and charge-regulating spheres, compared with the DLVO potential using a renormalized charge $Z^* = 663$. (c) Colloidal force $\beta F a$ for constant-charge and charge-regulating spheres, compared with the corresponding  DLVO force.}
\label{fig:twocol_pot}
\end{figure*}

To validate the finite-element Poisson-Boltzmann framework, we compare the resulting interaction potentials for a system consisting of two colloids with the DLVO potential. The DLVO potential was derived from Poisson's equation combined with the Boltzmann distribution to describe the interaction potential between two isolated charged colloids suspended in an electrolyte.\cite{Derjaguin1941, Verwey1948book} Since both the finite-element Poisson-Boltzmann calculations and the DLVO potential are based on the same underlying mean-field theory, they are expected to agree well for systems containing only two colloids---provided the surface charge is low enough to be in the linear-screening regime. For fixed charge, the DLVO potential is given by
\begin{equation}    
    \label{eq:DLVO}
    \beta U_{DLVO}(r) = Z^2 \lambda_B \left( \frac{\exp{(\kappa a)}}{1 + \kappa a} \right)^2 \frac{\exp{(-\kappa r)}}{r}, 
\end{equation}
where $r$ is the center-to-center distance between two colloids of charge valency $Z$. By taking the gradient of the potential $\mathbf{F}_{DLVO}=-\nabla U_{DLVO}$ we obtain the corresponding colloid force 
\begin{equation}    
    \label{eq:Force_DLVO}
    \beta F_{DLVO}(r) =  \hspace{-1mm}Z^2 \lambda_B \left( \frac{\exp{(\kappa a)}}{1 + \kappa a} \right)^2\hspace{-1mm} \left(\frac{1}{r} + \kappa\right)\frac{\exp{(-\kappa r)}}{r}.
\end{equation}

In Fig. \ref{fig:twocol_pot}, we examine the interaction between two isolated colloids. The parameters are chosen to match the experimental system studied by Monovoukas and Gast, who observed a gas-crystal phase separation under these conditions.\cite{Monovoukas1989} The colloids have radius $a = 66.7~\text{nm}$ and carry a fixed charge of valency $Z = 1200$ such that $\sigma=0.021\,\text{nm}^{-2}$. The Bjerrum length for water at room temperature is $\lambda_B = 0.72~\text{nm}$, corresponding to an effective temperature of $a/\lambda_B = 92.7$, and the screening length is $\kappa^{-1} = 170.1~\text{nm}$, yielding $\kappa a = 0.392$. Within linear screening theory, the resulting zeta-potential exceeds $9\,k_BT/e$ for these parameters, such that charge-renormalisation ($Z\rightarrow Z^*$) is to be expected for a quantitative comparison between the DLVO expressions of Eqs.(\ref{eq:DLVO}) and (\ref{eq:Force_DLVO}) with our numerical fully nonlinear Poisson-Boltzmann results.  
In addition, we consider charge-regulating colloids, for which the  surface charge density is determined using Eq. (\ref{eq:alpha_Bies}). To ensure that an isolated colloid has a total charge valency of $Z = 1200$, we set the number of dissociation sites to $M = 10^5$, and impose $\text{pH} = 7.0$ and a dissociation constant $\text{p}K = 6.23$.

In Fig. \ref{fig:twocol_pot}(a), we first plot the surface charge density $\sigma(\mathbf{r})$ of two charge-regulating colloids. 
At a relatively large separation $R = 4a$, the surface charge density is nearly constant over the entire colloid surface, with a value at around $\sigma  = 0.021~ \text{nm}^{-2}$, indeed very close to that of the constant-charge case. In contrast, at close-contact separation $R = 2a$, the charge distribution becomes much more heterogeneous, such that the charge density decreases to nearly $\sigma  \simeq 0 \ e/\text{nm}^2$ at the points of closest contact. This shows explicitly how the electrostatic  potential of a second colloid influences the surface charge through charge regulation.

The pairwise grand potentials $\beta\Omega^{(2)}$, as determined using Eqs.~(\ref{eq:grand-potential}) and (\ref{eq:grand-potential-constant-charge}), are plotted in Fig. \ref{fig:twocol_pot}(b) and compared to the DLVO potential of Eq. (\ref{eq:DLVO}). Here, we use the renormalized charge valency $Z^* = 663$ rather than the bare charge $Z = 1200$, as obtained using the cell model theory of Alexander \textit{et al.}, evaluated  at an effective packing fraction of $\eta = 0.00001$.\cite{Alexander1984, Trizac2003} The two-body contribution  $\beta \Omega^{(2)}$ obtained from the finite-element  calculations agrees reasonably well with the DLVO results, especially for asymptotically large particle-particle distances for which $Z^*$ is designed. For smaller distances, the DLVO potential is slightly more repulsive than both our results for $\Omega^{(2)}$. At large separations, the potentials for constant-charge colloids and charge-regulating colloids, blue and red dots, respectively, nearly coincide. However, for shorter separations, the two potentials deviate from each other, with charge-regulating colloids being less repulsive. This behavior is consistent with the lower surface charge at close distances due to charge regulation. 
In Fig. \ref{fig:twocol_pot}(c), we present the  corresponding (dimensionless) inter-colloidal forces $\beta Fa$. We find a similarly good agreement between the finite-element Poisson-Boltzmann  forces and the colloidal forces predicted by DLVO theory. As for the potentials, we find that the forces for constant-charge and charge-regulating colloids coincide at large separations, but deviate at short distances, where charge regulation leads to  a weaker repulsive force.

\subsection{Higher-body interactions}\label{sec:higher-body-interactions}

\begin{figure}
\centering
\includegraphics[width=0.999\linewidth]{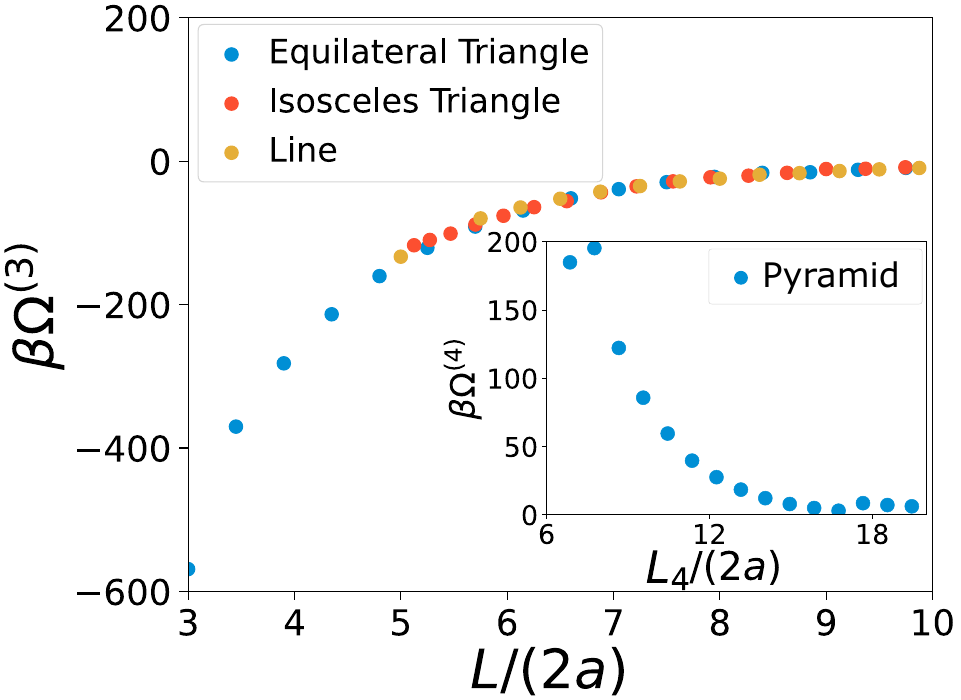}
\caption{\textbf{Three-body grand potential of charge-regulating colloids.} The three-body contribution $\beta \Omega^{(3)}$ for charge-regulating colloids at effective temperature $a/\lambda_B =92.7 $, screening parameter $\kappa a  = 0.392$, and charge valency $Z = 1200$. Charge regulation is implemented with  $M = 10^5$, $\text{pH} = 7.0$ and $\text{p}K = 6.23$. The potential is plotted as a function of the perimeter length $L$ of the triangle formed by three colloids, where we compare three colloid configurations. Inset: four-body contribution $\beta \Omega^{(4)}$ for an equilateral pyramid configuration of four colloids, as a  function of the total edge length $L_4$ of the pyramid. }
\label{fig:threecol_pot_shape}
\end{figure}

A central result of Russ {\em et al.} was the identification of an attractive three-body potential that depends (to a very good approximation) only on the total perimeter of the triangle formed by the three colloids.\cite{Russ2002} 
For the charge-regulating system with parameters discussed above, we plot in Fig. ~\ref{fig:threecol_pot_shape} the three-body potential $\beta \Omega^{(3)}$ as a function of the perimeter length  $L=R_{12}+R_{13}+R_{23}$ of the triangle for three representative configurations, a linear arrangement,  an isosceles triangle, and an equilateral triangle. 
As was the case for constant-charge colloids, we find for these charge-regulating colloids that the three-body interaction is attractive and depends essentially only on  the  perimeter $L$, with no measurable dependence on the specific particle configuration. 

In the inset of Fig. \ref{fig:threecol_pot_shape}, we also plot  the four-body contribution $\beta \Omega^{(4)}$ for the charge-regulating system, obtained  by placing four colloids in an equilateral pyramid configuration, with total edge length  $L_4$. In contrast to  the attractive three-body interaction, the four-body contribution is positive for all values of $L_4$, and therefore leads to a more repulsive colloid interaction. Interestingly, this behavior differs qualitatively from the results of Russ {\em et al.}, who found an attractive four-body contribution for constant-charge particles.\cite{Russ2002}  In our PB calculations of constant charge colloids we observe a repulsive four-body contribution, which is in qualitative agreement with our charge-regulation results, although it is in disagreement with the literature results of Russ {\em et al.} \cite{Russ2002}. However, the repulsive potential  agrees qualitatively with direct experimental measurements of the four-body potential.\cite{Dobnikar2004_four} 
The presence of this repulsive four-body contribution counteracts the attractive three-body interaction, which complicates the assessment of whether cohesive many-body effects can drive phase separation in colloidal suspensions. 

\subsection{Machine-learned potentials}\label{sec:ML-potentials}

By constructing interaction potentials using a machine-learning approach, we aim to capture the full set of many-body contributions to colloidal interactions. In the following section, we focus on a parameter regime where constant-charge colloids have valency $Z = 100$ rather than  $Z = 1200$.  By making Eqs. (\ref{eq:PB}) and (\ref{eq:PB-BC}) dimensionless, as was done by Russ {\em et al.},\cite{Russ2002} the Poisson-Boltzmann equation  depends on only two dimensionless parameters: the screening parameter $\kappa a$ and $Z \lambda_B/a$. Consequently, the system studied by Monovoukas and Gast, with colloids of valency $Z = 1200$ and effective temperature $a/\lambda_B = 92.7$, resembles the system with $Z = 100$ and effective temperature $a/\lambda_B = 7.7$, since both share the same value $Z \lambda_B /a  =12.94 $. To ensure that our charge-regulating colloids in isolation also have charge valency $Z=100$, we consider $M = 10^4$ charging sites on the colloid surface and impose $\text{p}K = 6.3$ and $\text{pH} = 7.0$.
However, although the Poisson-Boltzmann equation is identical for both systems, the resulting minimum of the grand potential (and therefore the effective interactions) is not. In particular, the grand potential depends not only on $\kappa a$ and $Z\lambda_B/a$, but also explicitly on the colloid charge $Z$. 
Nevertheless, we choose  the relatively low charge valency $Z = 100$ for our parameter sweep in order to enable direct comparisons with  detailed primitive model (PM) simulations. Such simulations become computationally impractical for highly charged colloids due to the large number of neutralizing counterions required.

\begin{figure}
\centering
\includegraphics[width=0.9999\linewidth]{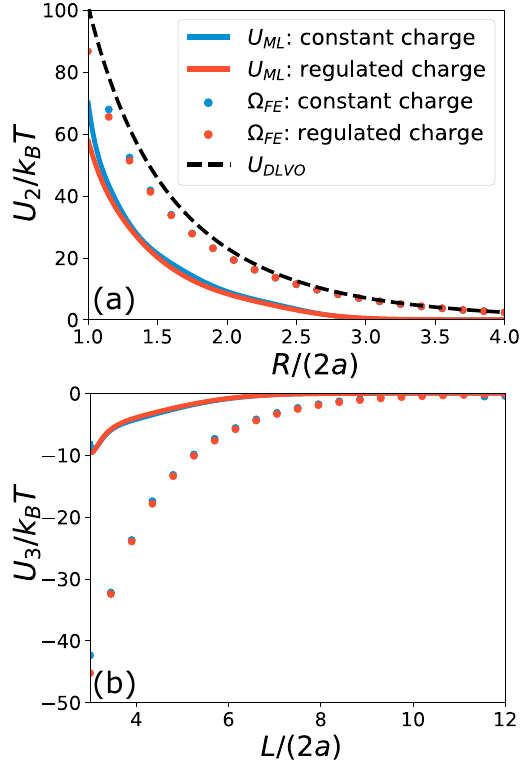}
\caption{\textbf{ML potential trained on Poisson-Boltzmann  calculations of 13 charged colloids} at effective temperature $a/\lambda_B =7.7 $, screening parameter $\kappa a  = 0.392$, and charge valency $Z = 100$. For charge-regulating colloids, the isolated valency $Z=100$ is obtained  using $M = 10^4$ charging sites at $\text{p}K = 6.3$ and $\text{pH} = 7.0$. (a) Two-body component of the ML potential $\beta U^{(2)}_{ML}$, compared to the potential $\beta \Omega^{(2)}$ obtained from PB calculations for  two colloids. For $U_{DLVO}$, a renormalized charge  $Z^* = 55$ is used. (b) Three-body component of the ML potential $\beta U^{(3)}_{ML}$, compared to the potential $\beta \Omega^{(3)}$ obtained from PB calculations for  three colloids. }
\label{fig:ML_PB_comparison}
\end{figure}

In Fig. \ref{fig:ML_PB_comparison}, we compare the ML potential $\beta U_{ML}$, trained on forces obtained from finite-element Poisson-Boltzmann calculations on clusters of 13 colloids, to the grand potential $\beta \Omega$ computed from finite-element Poisson-Boltzmann calculations of two isolated colloids. 
The ML potential is obtained using the procedure introduced in Section \ref{sec:ML-procedure}, while the grand potential is computed  as described in Section \ref{sec:PB-theory}. We present results for both constant-charge and  charge-regulating colloids. 
The two-body contribution is shown  in  Fig. \ref{fig:ML_PB_comparison}(a), where we compare the ML potential $\beta U^{(2)}_{ML}$ to the DLVO potential $\beta U_{DLVO}$ and to  the two-body potential $\beta \Omega^{(2)}$  obtained from finite-element Poisson-Boltzmann calculations of two isolated colloids. 
Note that we use a cutoff radius of $R_c =8.0a$ for the two-body and three-body symmetry functions. We verified that increasing the cutoff  radius to $R_c = 16.0$ yields   qualitatively similar ML potentials. 
As in Section \ref{sec:twocol}, we use a renormalized charge  $Z^*=55$ for the DLVO potential instead of the bare charge $Z$, as obtained from a Poisson-Boltzmann cell model calculation at a colloid packing fraction of $\eta = 0.00001$.  
Fig. \ref{fig:ML_PB_comparison} shows that the potentials for constant-charge  and charge-regulating colloids are in close agreement, with the two-body potential for  constant-charge colloids being slightly more repulsive, consistent with our findings  in Section \ref{sec:twocol}. We also see that the two-body  component of the ML potential, $\beta U^{(2)}_{ML}$, is considerably less repulsive than the corresponding  two-colloid PB result, $\beta \Omega^{(2)}$, although both potentials are of the same order of magnitude.  
In contrast, the difference in the three-body contributions shown in Fig. \ref{fig:ML_PB_comparison}(b) is more substantial. The ML potential yields a significantly reduced attractive three-body contribution, $\beta U^{(3)}_{ML}$, compared to  direct PB calculations of three colloids, $\beta \Omega^{(3)}$. This indicates that higher-order interaction terms, which contain partially repulsive contributions, are now redistributed into  the effective two- and three-body components of the ML potential, leading to the observed discrepancy between the ML potentials and the two-body and three-body PB potentials. In particular, the weaker three-body  contribution to the ML  potential suggests that the four-body and higher-order  repulsions are effectively absorbed into the lower-order terms.  Finally,  Fig.~\ref{fig:ML_PB_comparison}(b) shows that the effect of charge regulation is minimal compared to the constant-charge case in  both the ML and PB 
potentials.

\begin{figure}[h]
\centering
\includegraphics[width=0.9999\linewidth]{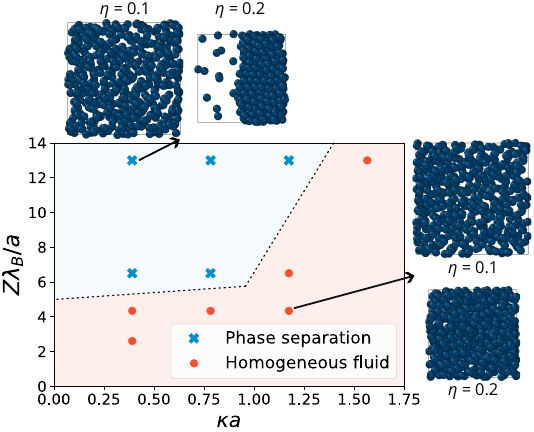}
\caption{\textbf{State diagram in the $ {Z\lambda_B/a}$ - ${\kappa a }$ representation}, showing the  regimes in which  gas-solid phase coexistence is observed (blue crosses) and those in  which homogeneous fluid states occur (orange dots). The results are based on   simulations using ML potentials generated for colloids with constant valency $Z = 100$, trained on finite-element Poisson-Boltzmann calculations of 13 colloids. Each point corresponds to  simulations performed with a separately trained ML potential. For two representative state points, snapshots of simulations are shown at packing fractions $\eta= 0.1$ and $\eta= 0.2$. The dashed line is a guide to the eye and should not be interpreted as a precise phase boundary.}
\label{fig:phase_plot_}
\end{figure}

Next, we apply the same training procedure, using finite-element Poisson-Boltzmann calculations of 13 constant-charge colloids with fixed valency $Z = 100$ as training data, while systematically  varying the effective temperature $a/\lambda_B$ and the screening parameter $\kappa a$. By performing simulations of $N = 500$ colloidal particles with the resulting ML  potentials, we identify the parameter regime in which phase separation emerges due  to attractive many-body interactions.   In the resulting state diagram, presented in Fig. \ref{fig:phase_plot_}, two regimes are observed: at low temperatures $a/\lambda_B$ and long screening lengths $1/\kappa a$, the attractive three-body contribution is strong enough to induce phase separation (blue crosses), whereas at higher temperatures and shorter screening lengths the system  remains homogeneous (orange dots). 
The parameter regime where one finds phase separation is in qualitative agreement with the estimate of Russ {\em et al.}, who  employed  an extension of  Van der Waals theory  to include three-body interactions in addition to pair interactions. However, their treatment only accounted for the Poisson-Boltzmann free-energy contributions up to third order, suggesting that the inclusion of higher-order interactions  does not qualitatively alter  the predicted  phase behavior.\cite{Russ2002}

The region of phase separation highlights the increasing importance of many-body interactions for longer screening lengths, where overlap between electric double layers becomes more pronounced. Similarly, we find that many-body interactions become more prominent for stronger electrostatic interactions between the colloids, corresponding to higher values of $Z\lambda_B/a$.  
We note, however, that phase separation is only observed at sufficiently high colloid densities (in particular only for packing fractions $\eta \gtrsim  0.2$), as shown in the snapshots of Fig. \ref{fig:phase_plot_}. Apparently, only in this regime does the three-body interaction component provide enough counterweight against the repulsive two-body interaction component to drive phase separation.  

\subsection{``Bulk'' machine-learned potentials}\label{sec:48colloids}
One should consider that the Poisson-Boltzmann calculations used as training data for the ML potential are not performed with periodic boundary conditions. Rather, the colloids are placed in the relatively small central region of a large spherical cell of radius $R=35a$, while a  boundary condition $\phi(R) = 0$ is imposed at the outer cell edge, as described in Eq. (\ref{eq:PB-BC}). This setup introduces a bias in the training data, since a large fraction of the 13 particles resides near the cluster boundary and is thus adjacent to the ``supernatant'' electrolyte rather than to other colloids in the dispersion. This is unlike the situation in bulk suspensions where  colloids are more homogeneously distributed. To assess the impact of this  effect  on the trained ML potentials, we  increase the number of particles in the training configurations and  exclude  training on forces acting on colloids located near the edge of the colloid clusters. 
For the state point considered in Fig. \ref{fig:ML_PB_comparison}, we therefore perform extended Poisson-Boltzmann calculations following the procedures described in Sections \ref{sec:PB-theory} and \ref{sec:ML-procedure}, but now  using 48 colloidal particles rather than 13, with effective packing fractions  $\eta \in[0.006, 0.65]$.  
The ML potential is then trained exclusively on the 13 colloids  closest to the cluster center, as illustrated in Fig. \ref{fig:workflow}(c). This  choice ensures that the  particles used for training are predominately  surrounded  by other colloids rather than by a supernatant electrolyte, thereby providing a closer approximation to a  bulk-like local environment. 
Importantly, the fitting protocol, truncation, and number of training particles are identical for the 13-particle potential and the 48-particle potential. The only difference is the local interaction environment represented in the respective training datasets. For the 48-particle potential, the training forces correspond to the 13 colloids located at the center of the 48-colloid cluster, whereas for the 13-particle potential, they correspond to colloids within the smaller cluster. Thus, the training dataset is deliberately shifted from configurations representative of colloids near a cluster edge to those representative of colloids embedded within a bulk-like environment.

\begin{figure}[h]
\centering
\includegraphics[width=0.9999\linewidth]{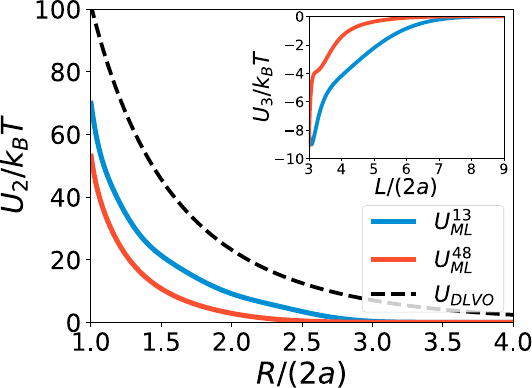}
\caption{\textbf{ML potential $\beta U^{48}_{ML}$ trained on Poisson-Boltzmann calculations of 48 charged colloids} at effective temperature $a/\lambda_B =7.7 $, screening parameter $\kappa a  = 0.392$, and charge valency $Z = 100$. For training,  only the  inner 13 ``bulk'' colloids are selected. A comparison is made to the potential trained on 13 isolated colloids $\beta U^{13}_{ML}$.  The two-body contribution of the ML potential $\beta U^{(2)}_{ML}$ is shown in the main panel, while the three-body contribution $\beta U^{(3)}_{ML}$ is shown  in the inset. For the DLVO potential, a  renormalized charge valency $Z^* = 55$ is used. }
\label{fig:13_48_comparison}
\end{figure}

This modified choice of training data   has a significant impact on the resulting potentials, as shown in Fig. \ref{fig:13_48_comparison}, where we compare our earlier potential trained on 13 isolated colloids ($U^{13}_{ML}$) to that trained on the 13 inner  ``bulk'' colloids from clusters of 48 colloids ($U^{48}_{ML}$). In both cases, we consider  colloids with charge valency $Z =100$,  at temperature $a/\lambda_B =  7.7$, and screening parameter $\kappa a = 0.392$. The two-body component of the bulk-trained ML potential is seen to be less repulsive than that of the  original ML model. 
However, this effect is negated by a reduction in the three-body attractions, since the attractive three-body contribution is significantly weaker for the ``bulk-trained'' ML potential. Molecular dynamics simulations using the bulk ML potential show  that the reduced three-body term is now insufficient to  provide  the cohesion required for like-charged phase separation, and  only homogeneous colloidal suspensions are observed in our simulations. 
Notably, broad phase coexistence is not observed even though the parameter set lies well within  the  region predicted to drive phase separation based on previous Poisson-Boltzmann studies \cite{Russ2002} and  our own  results based on ML potentials trained on Poisson-Boltzmann results  of 13 colloids (see Fig. \ref{fig:phase_plot_}). 
Thus, these results indicate that training on isolated clusters of colloids results in an overestimation of the role of  many-body attractions in bulk systems. 

In addition, we generated potentials for the system  described in Figs. \ref{fig:twocol_pot} and \ref{fig:threecol_pot_shape}, consisting of  colloids with valency $Z = 1200$ at temperature $a/\lambda_B = 92.7$ and screening parameter $\kappa a = 0.392$. These potentials exhibit   behavior analogous to those obtained for colloids with valency $Z =100$ and $a/\lambda_B = 7.7$, when training is performed on the forces acting on the inner 13 colloids of 48-particle  clusters. Compared to the potential trained on  forces from 13-colloid clusters, the attractive three-body contribution is  reduced, and no phase coexistence is observed in simulations using this ML potential. We present this potential in Appendix \ref{sec:app:bulkpotential}.

\subsection{Primitive model simulations}

The potentials trained on 13-colloid clusters $U^{13}_{ML}$, presented in Section \ref{sec:ML-potentials}, together with simulations performed using these potentials, suggest that  like-charged colloidal phase separation occurs at long screening lengths $1/(\kappa a)$ and low effective temperatures $a/\lambda_B$. However, we proceeded in Section \ref{sec:48colloids} by using the forces acting on the inner 13 colloids within  48-colloid clusters to generate the potential $U^{48}_{ML}$, in order to more accurately reproduce bulk-like conditions in the training data. Notably, no phase separation occurred in simulations employing this potential. 
We assess the accuracy of these potentials and their predictions regarding like-charge  phase separation through additional simulations in the primitive model (PM).  Unlike the Poisson-Boltzmann description, the primitive model treats ions explicitly rather than integrating out the ion degrees of freedom. 
In the primitive model, all particles interact via a combination of an hard-sphere potential and a Coulomb interaction 
\begin{equation}
    \hspace{-2mm} U_{PM}(r_{kl}) =\begin{cases}\infty,& \  \hspace{-2mm} r_{kl} \leq  2a_{kl};
        \\  k_B T \frac{Q_k Q_l \lambda_B}{r_{kl}} , & \  \hspace{-2mm}  r_{kl} >  2a_{kl}, 
    \end{cases}
    \label{eq:PM_pot}
\end{equation}
where $Q_k$ denotes the charge valency of particle $k$, and $r_{kl} = |\mathbf{r}_k - \mathbf{r}_l|$ is the distance between particle $k$ and $l$. In our PM simulations, three particle species are present: colloids of  radius $a$ and  charge valency $Z$, and coions and counterions of radius $a_i$ with valencies $-1$ and $+1$, respectively. The contact distance $a_{kl} = (a_k + a_l)/2$ in Eq. (\ref{eq:PM_pot}) follows directly from the radii of the interacting particles $k$ and $l$. 
We perform the PM simulations using molecular dynamics, in which  the infinitely steep repulsive potential described by Eq. (\ref{eq:PM_pot}) cannot be implemented directly. Hence, we approximate the short-ranged hard-sphere repulsion using the Weeks-Chandler-Andersen (WCA) potential, which for two particles $k$ and $l$ is given by

\begin{equation}
    \hspace{-2mm} U_{WCA}(r_{kl}) =\begin{cases}4\overline{\varepsilon}\left[\left(\frac{2a_{kl}}{r_{kl}}\right)^{12} \hspace{-2mm} - \left(\frac{2a_{kl}}{r_{kl}}\right)^{6} \right] + \overline{\varepsilon},& \  \hspace{-2mm} r_{kl} \leq 2^{7/6} a_{kl};
        \\ 0, & \  \hspace{-2mm} r_{kl} > 2^{7/6} a_{kl}. 
    \end{cases}
\end{equation}
We use a value of $\overline{\varepsilon} =40 k_BT $, which provides a good approximation  to  hard-sphere interactions.\cite{filion-2011}

\begin{figure*}[ht]
\centering
\includegraphics[width=0.999\linewidth]{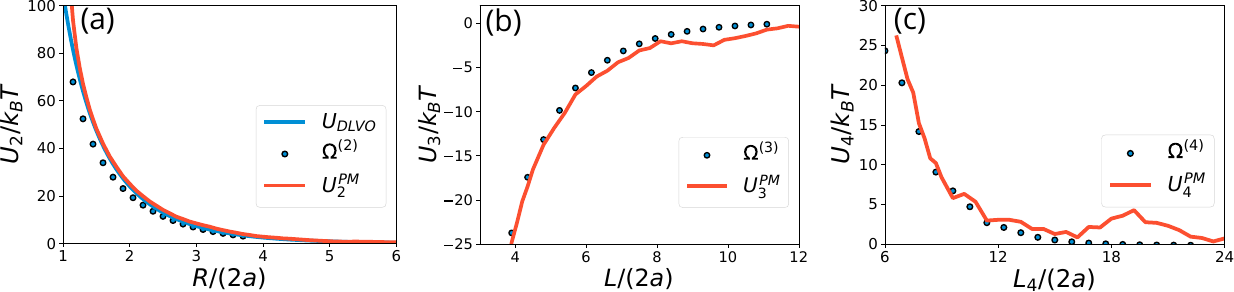}
\caption{\textbf{Potential of Mean Force (PMF) obtained from PM simulations} on (a) two isolated colloids, $U_{2}^{PM}$,  as a function of separation distance $R$, (b)  of three isolated colloids, $U_{3}^{PM}$, in an equilateral triangle  as a function of the total triangle perimeter length $L$, and of four isolated colloids, $U_{4}^{PM}$, in an equilateral tetrahedron  as a function of the total tetrahedron perimeter length $L_4$}. Each colloid carries a constant charge valency  $Z = 100$, at temperature $a/\lambda_B = 7.7$, and screening parameter $\kappa a = 0.392$. For the DLVO potential $U_{DLVO}$, we use a renormalized colloid charge valency $Z^* = 55$. Using a Bjerrum length of $\lambda_B=0.72$ nm for water at room temperature, we obtain an ion radius of  $a_i=0.27$ nm, which is comparable to the radius of a hydrated Na$^+$ ion. The grand potential $\Omega$ was obtained from finite-element Poisson-Boltzmann calculations on isolated colloid pairs and triplets, as described in Section \ref{sec:grandpot}. 
\label{fig:PMF_Z100}
\end{figure*}
We perform a range of PM simulations to investigate if phase separation would arise at  the same parameter set,  using both  canonical and  semi-grand canonical ensemble simulations. Furthermore, we perform  simulations over a range of colloid packing fractions, and consider both  initially homogeneous configurations, and configurations prepared as  a crystal adjacent to an  electrolyte. In all cases, no signatures of stable phase coexistence are observed, which agrees with Ref. \citenum{Hynninen2005}, where the cohesive effects of many-body interactions were found to be minimal in PM simulations.
This is consistent with earlier results of Ref. \citenum{terRele2026}, where ML potentials  trained on PM simulations also did not exhibit phase separation in the relevant parameter regime. 
Additionally, the absence of phase separation in PM simulations agrees with the potentials generated in Section \ref{sec:48colloids}, which did not support  phase coexistence either. 

To further clarify the discrepancy between the simulations based on ML potentials trained on Poisson-Boltzmann calculations and on PM simulations, we focus on a representative state point for which the two-body and three-body Poisson-Boltzmann  potentials predict phase separation. Specifically, we consider the system shown in Fig. \ref{fig:ML_PB_comparison}, at temperature $a/\lambda_B =  7.7$, screening parameter $\kappa a = 0.392$, and constant  charge valency $Z = 100$. This relatively low valency enables us to perform direct PM simulations to determine the many-body interaction potential.

To accurately determine the PM interaction potential between two colloids suspended in an electrolyte, we compute the potential of mean force (PMF) $U_{2}^{PM}(R)$ as a function of the interparticle separation $R$. We obtain the PMF from PM simulations by fixing two colloids at a series of separations $R$ and measuring the corresponding ion-averaged force ${\bf F}^{PM}_i(R)$ acting on the colloids. The simulations are performed in a cubic box of side length $L = 60a$, with colloid separations ranging from $R = 2.2 a$ to $R = 12 a$. We consider the system described in Fig. \ref{fig:13_48_comparison}, consisting of constant-charge colloids with valency $Z = 100$, at an effective temperature $a/\lambda_B = 7.7$, and screening length $\kappa a = 0.392$. By integrating the force ${\bf F}^{PM}_i(R)$ over the separation distance $R'$, we obtain the potential of mean force
\begin{equation}\label{U2PM}
    U_{2}^{PM}(R) = \int_R^\infty dR' \frac{\mathbf{F}^{PM}_i(R') \cdot \mathbf{R}_{ji}}{R'},
\end{equation}
where $\mathbf{R}_{ji}={\bf R}_i-{\bf R}_j$ denotes the vector connecting colloid $j$ and $i$. 
In addition, we  perform simulations of three colloids (labeled $i,j,k$) arranged in an equilateral triangle of fixed side length $R$ (and  perimeter $L=3R$), and we again measure the ion-averaged forces ${\bf F}^{PM}_i(R)$ acting on each colloid. By integrating this force we obtain the three-body PMF 
\begin{eqnarray}
     U_{3}^{PM}(R) =  \hspace{-18mm} &\\ 
    &3\left[\frac{1}{3} \int_R^\infty dR' \frac{\mathbf{F}^{PM}_i(R') \cdot \mathbf{R}_{ji} + \mathbf{F}^{PM}_i(R') \cdot \mathbf{R}_{ki}}{R'} \nonumber  - U^{PM}_2(R)\right],
    \label{U3PM}
\end{eqnarray}
where the factor $1/3$ is included to account for the triangular configuration of the colloids.\cite{Zhang2016, terRele2025}  The three-body contribution $U_{3}^{PM}(R)$ to the potential of mean force  describes the additional interaction  between groups of three colloids that cannot be explained by  two-body contributions. 
Furthermore, we perform simulations of four colloids (labeled $i,j,k,l$) arranged in an tetrahedral configurations with edge length $R$. The tetrahedron consists of four equilateral triangular faces, corresponding to configurations that  we simulated before.  By integrating the ion-averaged forces acting on each colloid, we obtain the four-body PMF
\begin{align}
     \hspace{-7mm}U_{4}^{PM}(R) =&  6\Bigg[\frac{1}{6} \int_R^\infty dR' \frac{\sum_{m\neq i} \mathbf{F}^{PM}_i(R') \cdot \mathbf{R}_{mi}}{R'}\nonumber
     \\ &- U^{PM}_2(R)\Bigg] - 4U_3^{PM}(R),
    \label{U4PM}
\end{align}
where the sum runs over all colloids except  the colloid with label $i$.
The four-body contribution $U_{4}^{PM}(R)$ represents the part of the interaction that cannot be decomposed into two- and three-body potential contributions.

In Fig.~\ref{fig:PMF_Z100}(a), we compare the two-body potential of mean force  $U_{2}^{PM}$,  obtained from PM simulations of two colloids with the DLVO potential $ U_{DLVO}$, and the two-body grand-potential contribution $\Omega^{(2)}$, obtained from finite-element Poisson-Boltzmann calculations of a colloid pair. In  Fig.~\ref{fig:PMF_Z100}(b), we compare the three-body potential of mean force contribution $U_{3}^{PM}$,  obtained from PM simulations of three  colloids arranged in an equilateral configuration, with the three-body contribution  $ \Omega^{(3)}$ of the grand potential obtained from finite-element Poisson-Boltzmann calculations of a colloid triplet. 

Furthermore, in Fig.~\ref{fig:PMF_Z100}(c), we compare the four-body potential of mean force  $U_{4}^{PM}$,  obtained from PM simulations of colloids arranged in a tetrahedral configuration, with the corresponding four-body grand-potential contribution $ \Omega^{(4)}$. 
The finite-element Poisson-Boltzmann calculations are in good agreement with the PM simulations, further confirming  the applicability of Poisson-Boltzmann theory  in this parameter regime.      
We find excellent quantitative agreement for the two- and three-body contributions and qualitative agreement for the four-body contribution. Importantly, both the primitive-model PMF and the finite-element Poisson–Boltzmann calculations yield a repulsive four-body contribution. 
We expect that  higher-order terms  within the primitive model will predominately be repulsive, consistent with trends observed in our Poisson-Boltzmann calculations. 
This is in line with  the absence of phase coexistence observed in like-charged colloids in PM simulations at the state point shown in Fig. \ref{fig:PMF_Z100}.

Note that in Fig.~\ref{fig:PMF_Z100}, we show the three- and four-body potential contributions, that we explicitly extracted from simulations and calculations of isolated clusters of three and four colloids, respectively. This results in excellent agreement between the three- and four-body PMFs, $U_{3}^{PM}$ and $U_{4}^{PM}$, and the corresponding contributions to the grand potential, $ \Omega^{(3)}$ and $ \Omega^{(4)}$. In contrast,  the three-body ML potential component $U_{ML}^{(3)}$ in Fig. \ref{fig:ML_PB_comparison} is obtained by fitting to forces acting on particles within a many-particle environment. Hence, it  incorporates contributions from higher-order interactions, including the repulsive four-body contribution, resulting in substantial differences between $U_{ML}^{(3)}$ and the isolated three-body contribution $\Omega^{(3)}$.

\section{Conclusion}

In conclusion, we have employed finite-element Poisson-Boltzmann calculations to investigate the many-body  nature of the interactions between suspended charged colloids. To obtain a more realistic description, we further incorporate two additional ingredients into the model. 

First, we take into account charge regulation to introduce density dependence in the colloidal surface charge, providing a  more realistic description than the commonly used  constant-charge assumption. By computing two-, three-, and four-body interaction  potentials using Poisson-Boltzmann  calculations, we found that charge regulation slightly reduces  the short-range repulsion  compared to  constant-charge colloids, although the effect is  marginal. Moreover, we find that charge-regulating colloids interact via an attractive three-body potential that depends solely on the total perimeter of the triangle formed by each triplet of colloids, in agreement with previous results on constant-charge colloids. We thus find that this configurational independence persists also for charge-regulating spheres. 

Second, we employed a machine-learning framework to fit the many-body colloidal interactions. This approach enables a more accurate assessment of whether  many-body potentials can provide sufficient cohesive energy to drive gas-liquid or gas-solid  phase separation.  We observed from the ML potentials that only considering pair and triplet interactions does not suffice when studying suspensions of colloids with large screening lengths $\kappa^{-1}$. In particular, the ML three-body potential that takes into account higher-order contributions was significantly less attractive than the Poisson-Boltzmann  potential obtained from  isolated three-colloid configurations, indicating that the inclusion of  higher-order terms  weakens the overall cohesive nature of the many-body potential. 
When larger clusters of colloids were included in the ML training set, thereby increasing the influence of higher-order interactions, the inferred three-body potential became even less attractive.
As these higher-order interactions gained importance in the ML training data, the resulting ML potentials no longer induced broad phase coexistence. This indicates that earlier studies likely overestimated  the attractive nature of many-body interactions, as the potentials were derived through calculations on only a limited number of isolated particles.

Additionally, no phase separation was observed in PM simulations for any set of parameters that we considered, resembling colloids dispersed in an aqueous 1:1 electrolyte. Finally, we computed the potential of mean force of pairs, triplets and quartets of colloids using primitive model simulations. 
These results show  excellent agreement with the potentials obtained from Poisson-Boltzmann calculations, thereby supporting the validity of the Poisson-Boltzmann  approach for determining many-body interactions. 

In  conclusion, we have constructed many-body interaction potentials for charged colloids using a range of complementary approaches. Finite-element Poisson-Boltzmann calculations reveal an attractive three-body potential that could provide, in principle,  the cohesive energy required  for broad phase separation. However, we found that upon  incorporating  higher-order  contributions, this attraction is progressively diminished,  and the tendency toward gas-liquid or gas-solid  phase separation disappears. This picture is corroborated by PM simulations, in which no broad phase separation was  observed in the presently studied parameter regime of aqueous colloidal dispersions in 1:1 electrolytes. 
Therefore, we question the hypothesis that attractive many-body interactions are  at the origin of the anomalous behavior of like-charge attractions observed in experiments. While three-body contributions are indeed attractive, higher-order terms generally are not, and their combined effect is insufficient to drive gas-liquid phase separation within  Poisson-Boltzmann  theory. We therefore do not expect broad bulk gas-liquid or gas-solid phase separation in suspensions of like-charged colloids in aqueous dispersions to arise solely from  attractive many-body interactions. The experimental observations of like-charge attractions in these dispersions require explanations beyond the primitive model in equilibrium.

\begin{acknowledgments}
The authors thank Tim Veenstra for many useful discussions, and Nena Slaats for her sharp attention to an early version of this manuscript. T.t.R. and M.D. acknowledge funding from the European Research Council (ERC) under the European Union's Horizon 2020 research and innovation programme (Grant agreement No. ERC-2019-ADG 884902 SoftML).
\end{acknowledgments}

\section*{Conflict of Interest}
The authors have to conflicts to disclose.

\section*{Author Contribution}
\textbf{Thijs ter Rele}: Conceptualization (equal); Data Curation (lead); Formal Analysis (lead); Investigation (lead); Methodology (equal); Software (lead); Visualization (lead); Writing - original draft (lead); Writing - review \& editing (equal). 
\textbf{Ren\'{e} van Roij}: Conceptualization (equal);  Investigation (supporting); Methodology (equal); Visualization (supporting);   Writing - review \& editing (equal). 
\textbf{Marjolein Dijkstra}: Conceptualization (equal), Funding acquisition (lead); Investigation (supporting); Methodology (equal); Supervision (lead);  Visualization (supporting);  Writing - review \& editing (equal). 

\section*{Data Availibility Statement}
Data will be made available on request. Simulation and analysis code supporting the results presented in this manuscript is provided as supplementary material.

\bibliography{main}

@article{Derjaguin1941,
  doi = {10.1016/0079-6816(93)90013-l},
  url = {https://doi.org/10.1016/0079-6816(93)90013-l},
  year = {1941},
  volume = {14},
  pages = {633--662},
  author = {B. Derjaguin and L. Landau},
  title = {Theory of the stability of strongly charged lyophobic sols and of the adhesion of strongly charged particles in solutions of electrolytes},
  journal = {Acta Physicochimica U.R.S.S.}
}

@book{Verwey1948book,
	author = {Verwey, E. J. W. and Overbeek, J. Th. G.},
	month = {1},
    publisher = {Elsevier Publishing Company},
	title = {{Theory of the stability of lyophobic colloids: the interaction of sol particles having and electric double layer}},
	year = {1948},
	url = {},
}

@article{Tata1992,
  doi = {10.1103/physrevlett.69.3778},
  url = {https://doi.org/10.1103/physrevlett.69.3778},
  year = {1992},
  month = dec,
  publisher = {American Physical Society ({APS})},
  volume = {69},
  number = {26},
  pages = {3778--3781},
  author = {B. V. R. Tata and M. Rajalakshmi and Akhilesh K. Arora},
  title = {Vapor-liquid condensation in charged colloidal suspensions},
  journal = {Physical Review Letters}
}

@article{Palberg1994_tata,
  title = {Comment on ‘“Vapor-liquid condensenation in charged colloidal suspensions”’},
  volume = {72},
  ISSN = {0031-9007},
  url = {http://dx.doi.org/10.1103/PhysRevLett.72.786},
  DOI = {10.1103/physrevlett.72.786},
  number = {5},
  journal = {Physical Review Letters},
  publisher = {American Physical Society (APS)},
  author = {Palberg,  T. and W\"{u}rth,  M.},
  year = {1994},
  month = jan,
  pages = {786--786}
}

@article{Kepler1994,
  title = {Attractive potential between confined colloids at low ionic strength},
  volume = {73},
  ISSN = {0031-9007},
  url = {http://dx.doi.org/10.1103/PhysRevLett.73.356},
  DOI = {10.1103/physrevlett.73.356},
  number = {2},
  journal = {Physical Review Letters},
  publisher = {American Physical Society (APS)},
  author = {Kepler,  Grace Martinelli and Fraden,  Seth},
  year = {1994},
  month = jul,
  pages = {356--359}
}

@article{Ito1994,
  doi = {10.1126/science.263.5143.66},
  url = {https://doi.org/10.1126/science.263.5143.66},
  year = {1994},
  month = jan,
  publisher = {American Association for the Advancement of Science ({AAAS})},
  volume = {263},
  number = {5143},
  pages = {66--68},
  author = {Kensaku Ito and Hiroshi Yoshida and Norio Ise},
  title = {Void Structure in Colloidal Dispersions},
  journal = {Science}
}

@article{Larsen1997,
  doi = {10.1038/385230a0},
  url = {https://doi.org/10.1038/385230a0},
  year = {1997},
  month = jan,
  publisher = {Springer Science and Business Media {LLC}},
  volume = {385},
  number = {6613},
  pages = {230--233},
  author = {Amy E. Larsen and David G. Grier},
  title = {Like-charge attractions in metastable colloidal crystallites},
  journal = {Nature}
}

@article{Squires2000_brenner,
  title = {Like-Charge Attraction and Hydrodynamic Interaction},
  volume = {85},
  ISSN = {1079-7114},
  url = {http://dx.doi.org/10.1103/PhysRevLett.85.4976},
  DOI = {10.1103/physrevlett.85.4976},
  number = {23},
  journal = {Physical Review Letters},
  publisher = {American Physical Society (APS)},
  author = {Squires,  Todd M. and Brenner,  Michael P.},
  year = {2000},
  month = dec,
  pages = {4976--4979}
}

@article{Wang2024,
  title = {A charge-dependent long-ranged force drives tailored assembly of matter in solution},
  volume = {19},
  ISSN = {1748-3395},
  url = {http://dx.doi.org/10.1038/s41565-024-01621-5},
  DOI = {10.1038/s41565-024-01621-5},
  number = {4},
  journal = {Nature Nanotechnology},
  publisher = {Springer Science and Business Media LLC},
  author = {Wang,  Sida and Walker-Gibbons,  Rowan and Watkins,  Bethany and Flynn,  Melissa and Krishnan,  Madhavi},
  year = {2024},
  month = mar,
  pages ={485--493}
}

@article{vanRoij1997,
  title = {Van der Waals–Like Instability in Suspensions of Mutually Repelling Charged Colloids},
  volume = {79},
  ISSN = {1079-7114},
  url = {http://dx.doi.org/10.1103/PhysRevLett.79.3082},
  DOI = {10.1103/physrevlett.79.3082},
  number = {16},
  journal = {Physical Review Letters},
  publisher = {American Physical Society (APS)},
  author = {van Roij,  René and Hansen,  Jean-Pierre},
  year = {1997},
  month = oct,
  pages = {3082--3085}
}

@article{vanRoij1999,
  doi = {10.1103/physreve.59.2010},
  url = {https://doi.org/10.1103/physreve.59.2010},
  year = {1999},
  month = feb,
  publisher = {American Physical Society ({APS})},
  volume = {59},
  number = {2},
  pages = {2010--2025},
  author = {Ren{\'{e}} van Roij and Marjolein Dijkstra and Jean-Pierre Hansen},
  title = {Phase diagram of charge-stabilized colloidal suspensions: van der Waals instability without attractive forces},
  journal = {Physical Review E}
}

@article{Russ2002,
  title = {Three-body forces between charged colloidal particles},
  volume = {66},
  ISSN = {1095-3787},
  url = {http://dx.doi.org/10.1103/PhysRevE.66.011402},
  DOI = {10.1103/physreve.66.011402},
  number = {1},
  journal = {Physical Review E},
  publisher = {American Physical Society (APS)},
  author = {Russ,  C. and von Gr\"{u}nberg,  H. H. and Dijkstra,  M. and van Roij,  R.},
  year = {2002},
  month = jul,
  pages = {011402}
}

@article{Zoetekouw2006,
  doi = {10.1103/physreve.73.021403},
  url = {https://doi.org/10.1103/physreve.73.021403},
  year = {2006},
  month = feb,
  publisher = {American Physical Society ({APS})},
  volume = {73},
  number = {2},
  author = {Bas Zoetekouw and Ren{\'{e}} van Roij},
  title = {Volume terms for charged colloids: A grand-canonical treatment},
  journal = {Physical Review E},
  pages = {021403}
}

@article{Warren2000,
  title = {A theory of void formation in charge-stabilized colloidal suspensions at low ionic strength},
  volume = {112},
  ISSN = {1089-7690},
  url = {http://dx.doi.org/10.1063/1.481024},
  DOI = {10.1063/1.481024},
  number = {10},
  journal = {The Journal of Chemical Physics},
  publisher = {AIP Publishing},
  author = {Warren,  Patrick B.},
  year = {2000},
  month = mar,
  pages = {4683 -- 4698}
}

@article{Zoetekouw2006prl,
  title = {Nonlinear Screening and Gas-Liquid Separation in Suspensions of Charged Colloids},
  volume = {97},
  ISSN = {1079-7114},
  url = {http://dx.doi.org/10.1103/PhysRevLett.97.258302},
  DOI = {10.1103/physrevlett.97.258302},
  number = {25},
  journal = {Physical Review Letters},
  publisher = {American Physical Society (APS)},
  author = {Zoetekouw,  Bas and van Roij,  René},
  year = {2006},
  month = dec,
  pages = {258302}
}

@article{Kubincov2020,
  doi = {10.1063/1.5141346},
  url = {https://doi.org/10.1063/1.5141346},
  year = {2020},
  month = mar,
  publisher = {{AIP} Publishing},
  volume = {152},
  number = {10},
  author = {Al{\v{z}}beta Kubincov{\'{a}} and Philippe H. H\"{u}nenberger and Madhavi Krishnan},
  title = {Interfacial solvation can explain attraction between like-charged objects in aqueous solution},
  journal = {The Journal of Chemical Physics}, 
  pages = {104713}
}

@article{terRele2025,
  title = {Machine learning many-body potentials for charged colloids in primitive 1:1 electrolytes},
  volume = {163},
  ISSN = {1089-7690},
  url = {http://dx.doi.org/10.1063/5.0291389},
  DOI = {10.1063/5.0291389},
  number = {16},
  journal = {The Journal of Chemical Physics},
  publisher = {AIP Publishing},
  author = {ter Rele,  Thijs and Campos-Villalobos,  Gerardo and van Roij,  René and Dijkstra,  Marjolein},
  year = {2025},
  month = oct,
  pages = {164106}
}

@article{Behler2007,
  doi = {10.1103/physrevlett.98.146401},
  url = {https://doi.org/10.1103/physrevlett.98.146401},
  year = {2007},
  month = apr,
  publisher = {American Physical Society ({APS})},
  volume = {98},
  number = {14},
  author = {J\"{o}rg Behler and Michele Parrinello},
  title = {Generalized Neural-Network Representation of High-Dimensional Potential-Energy Surfaces},
  journal = {Physical Review Letters},
  pages = {146401}
}

@article{Zhang2016,
  title = {Potential of mean force between like-charged nanoparticles: Many-body effect},
  volume = {6},
  ISSN = {2045-2322},
  url = {http://dx.doi.org/10.1038/srep23434},
  DOI = {10.1038/srep23434},
  number = {1},
  journal = {Scientific Reports},
  publisher = {Springer Science and Business Media LLC},
  author = {Zhang,  Xi and Zhang,  Jin-Si and Shi,  Ya-Zhou and Zhu,  Xiao-Long and Tan,  Zhi-Jie},
  year = {2016},
  month = mar,
  pages = {23434}
}

@article{Monovoukas1989,
  title = {The experimental phase diagram of charged colloidal suspensions},
  volume = {128},
  ISSN = {0021-9797},
  url = {http://dx.doi.org/10.1016/0021-9797(89)90368-8},
  DOI = {10.1016/0021-9797(89)90368-8},
  number = {2},
  journal = {Journal of Colloid and Interface Science},
  publisher = {Elsevier BV},
  author = {Monovoukas,  Yiannis and Gast,  Alice P},
  year = {1989},
  month = mar,
  pages = {533 -- 548}
}

@article{Singraber2019,
  title = {Library-Based LAMMPS Implementation of High-Dimensional Neural Network Potentials},
  volume = {15},
  ISSN = {1549 -- 9626},
  url = {http://dx.doi.org/10.1021/acs.jctc.8b00770},
  DOI = {10.1021/acs.jctc.8b00770},
  number = {3},
  journal = {Journal of Chemical Theory and Computation},
  publisher = {American Chemical Society (ACS)},
  author = {Singraber,  Andreas and Behler,  J\"{o}rg and Dellago,  Christoph},
  year = {2019},
  month = jan,
  pages = {1827 -- 1840}
}

@Article{LAMMPS,
  author = "A. P. Thompson and H. M. Aktulga and R. Berger and 
     D. S. Bolintineanu and W. M. Brown and P. S. Crozier and
     P. J. in 't Veld and A. Kohlmeyer and S. G. Moore and T. D. Nguyen and
     R. Shan and M. J. Stevens and J. Tranchida and C. Trott and S. J. Plimpton",
  title = "{LAMMPS} - a flexible simulation tool for
     particle-based materials modeling at the 
     atomic, meso, and continuum scales",
  journal = "Comp. Phys. Comm.",
  volume =  "271",
  pages =   "108171",
  year =    "2022",
  doi = "10.1016/j.cpc.2021.108171"
}

@book{Hockney1988,
	author    = {R.W Hockney and J.W Eastwood}, 
	title     = {Computer Simulation Using Particles},
	publisher = {CRC Press},
	year      = 1988,
	volume    = {},
	series    = {},
	address   = {},
	edition   = 1,
	month     = {},
	note      = {},
	isbn      = {9780367806934}
}

@article{ninham-1971,
	author = {Ninham, Barry W. and Parsegian, V. Adrian},
	journal = {Journal of Theoretical Biology},
	month = {6},
	number = {3},
	pages = {405--428},
	title = {{Electrostatic potential between surfaces bearing ionizable groups in ionic equilibrium with physiologic saline solution}},
	volume = {31},
	year = {1971},
	doi = {10.1016/0022-5193(71)90019-1},
	url = {https://doi.org/10.1016/0022-5193(71)90019-1},
}

@article{biesheuvel-2004A,
	author = {Biesheuvel, P.M. and Stuart, Martien A. Cohen},
	journal = {Langmuir},
	month = {3},
	number = {7},
	pages = {2785--2791},
	title = {{Electrostatic free energy of weakly charged macromolecules in solution and intermacromolecular complexes consisting of oppositely charged polymers}},
	volume = {20},
	year = {2004},
	doi = {10.1021/la036204l},
	url = {https://doi.org/10.1021/la036204l},
}

@article{Everts2016,
  title = {Tuning Colloid-Interface Interactions by Salt Partitioning},
  volume = {117},
  ISSN = {1079-7114},
  url = {http://dx.doi.org/10.1103/PhysRevLett.117.098002},
  DOI = {10.1103/physrevlett.117.098002},
  number = {9},
  journal = {Physical Review Letters},
  publisher = {American Physical Society (APS)},
  author = {Everts,  J. C. and Samin,  S. and van Roij,  R.},
  year = {2016},
  month = aug, 
  pages = {098002}
}

@article{Grunberg2001,
  title = {Charged colloids near interfaces},
  volume = {13},
  ISSN = {1361-648X},
  url = {http://dx.doi.org/10.1088/0953-8984/13/21/311},
  DOI = {10.1088/0953-8984/13/21/311},
  number = {21},
  journal = {Journal of Physics: Condensed Matter},
  publisher = {IOP Publishing},
  author = {Gr\"{u}nberg,  H H von and Mbamala,  E C},
  year = {2001},
  month = may,
  pages = {4801–4834}
}

@article{Dobnikar2003,
  title = {Effect of many-body interactions on the solid-liquid phase behavior of charge-stabilized colloidal suspensions},
  volume = {61},
  ISSN = {1286-4854},
  url = {http://dx.doi.org/10.1209/epl/i2003-00142-5},
  DOI = {10.1209/epl/i2003-00142-5},
  number = {5},
  journal = {Europhysics Letters (EPL)},
  publisher = {IOP Publishing},
  author = {Dobnikar,  J and Rzehak,  R and Gr\"{u}nberg,  H. H. von},
  year = {2003},
  month = mar,
  pages = {695–701}
}

@article{Dobnikar2004,
  title = {Three-body interactions in colloidal systems},
  volume = {69},
  ISSN = {1550-2376},
  url = {http://dx.doi.org/10.1103/PhysRevE.69.031402},
  DOI = {10.1103/physreve.69.031402},
  number = {3},
  journal = {Physical Review E},
  publisher = {American Physical Society (APS)},
  author = {Dobnikar,  Jure and Brunner,  Matthias and von Gr\"{u}nberg,  Hans-Hennig and Bechinger,  Clemens},
  year = {2004},
  month = mar,
  pages = {031402}
}

@inproceedings{Dobnikar2004_four,
  title = {Three- and four-body interactions in colloidal systems},
  volume = {5514},
  ISSN = {0277-786X},
  url = {http://dx.doi.org/10.1117/12.554610},
  DOI = {10.1117/12.554610},
  booktitle = {Optical Trapping and Optical Micromanipulation},
  publisher = {SPIE},
  author = {Dobnikar,  Jure and Brunner,  Matthias and Baumgartl,  Jorg and Bechinger,  Clemens and von Gruenberg,  Hans-Hennig},
  editor = {Dholakia,  Kishan and Spalding,  Gabriel C.},
  year = {2004},
  month = Oct,
  pages = {340}
}

@article{Brunner2004,
  title = {Direct Measurement of Three-Body Interactions amongst Charged Colloids},
  volume = {92},
  ISSN = {1079-7114},
  url = {http://dx.doi.org/10.1103/PhysRevLett.92.078301},
  DOI = {10.1103/physrevlett.92.078301},
  number = {7},
  journal = {Physical Review Letters},
  publisher = {American Physical Society (APS)},
  author = {Brunner,  Matthias and Dobnikar,  Jure and von Gr\"{u}nberg,  Hans-Hennig and Bechinger,  Clemens},
  year = {2004},
  month = feb ,
  pages = {078301}
}

@article{Hynninen2004,
  title = {Effect of three-body interactions on the phase behavior of charge-stabilized colloidal suspensions},
  volume = {69},
  ISSN = {1550-2376},
  url = {http://dx.doi.org/10.1103/PhysRevE.69.061407},
  DOI = {10.1103/physreve.69.061407},
  number = {6},
  journal = {Physical Review E},
  publisher = {American Physical Society (APS)},
  author = {Hynninen,  A.-P. and Dijkstra,  M. and van Roij,  R.},
  year = {2004},
  month = jun, 
  pages = {061407}
}

@article{Hynninen2005,
  title = {Melting line of charged colloids from primitive model simulations},
  volume = {123},
  ISSN = {1089-7690},
  url = {http://dx.doi.org/10.1063/1.2138693},
  DOI = {10.1063/1.2138693},
  number = {24},
  journal = {The Journal of Chemical Physics},
  publisher = {AIP Publishing},
  author = {Hynninen,  Antti-Pekka and Dijkstra,  Marjolein},
  year = {2005},
  month = dec, 
  pages = {244902}
}

@article{Merrill2009,
  title = {Many-Body Electrostatic Forces between Colloidal Particles at Vanishing Ionic Strength},
  volume = {103},
  ISSN = {1079-7114},
  url = {http://dx.doi.org/10.1103/PhysRevLett.103.138301},
  DOI = {10.1103/physrevlett.103.138301},
  number = {13},
  journal = {Physical Review Letters},
  publisher = {American Physical Society (APS)},
  author = {Merrill,  Jason W. and Sainis,  Sunil K. and Dufresne,  Eric R.},
  year = {2009},
  month = sep,
  pages = {138301}
}

@article{Alexander1984,
    author = {Alexander, S. and Chaikin, P. M. and Grant, P. and Morales, G. J. and Pincus, P. and Hone, D.},
    title = {Charge renormalization, osmotic pressure, and bulk modulus of colloidal crystals: Theory},
    journal = {The Journal of Chemical Physics},
    volume = {80},
    number = {11},
    pages = {5776-5781},
    year = {1984},
    month = {06},
    issn = {0021-9606},
    doi = {10.1063/1.446600},
    url = {https://doi.org/10.1063/1.446600},
}

@article{James1985,
title = {Numerical solution of the Poisson—Boltzmann equation},
journal = {Journal of Colloid and Interface Science},
volume = {107},
number = {1},
pages = {44-59},
year = {1985},
issn = {0021-9797},
doi = {https://doi.org/10.1016/0021-9797(85)90147-X},
url = {https://www.sciencedirect.com/science/article/pii/002197978590147X},
author = {A.E James and D.J.A Williams},
}

@article{Holst2000,
  title = {Adaptive multilevel finite element solution of the Poisson-Boltzmann equation I. Algorithms and examples},
  volume = {21},
  ISSN = {1096-987X},
  url = {http://dx.doi.org/10.1002/1096-987X(20001130)21:15<1319::AID-JCC1>3.0.CO;2-8},
  DOI = {10.1002/1096-987x(20001130)21:15<1319::aid-jcc1>3.0.co;2-8},
  number = {15},
  journal = {Journal of Computational Chemistry},
  publisher = {Wiley},
  author = {Holst,  M. and Baker,  N. and Wang,  F.},
  year = {2000},
  pages = {1319--1342}
}

@article{Holst2012,
  title = {Adaptive Finite Element Modeling Techniques for the Poisson-Boltzmann Equation},
  volume = {11},
  ISSN = {1991--7120},
  url = {http://dx.doi.org/10.4208/cicp.081009.130611a},
  DOI = {10.4208/cicp.081009.130611a},
  number = {1},
  journal = {Communications in Computational Physics},
  publisher = {Global Science Press},
  author = {Holst,  M. and McCammon,  J.A. and Yu,  Z. and Zhou,  Y.C. and Zhu,  Y.},
  year = {2012},
  month = jan,
  pages = {179--214}
}

@article{Schlaich2023,
  title = {Renormalized charge and dielectric effects in colloidal interactions: a numerical solution of the nonlinear Poisson–Boltzmann equation for unknown boundary conditions},
  volume = {46},
  ISSN = {1292--895X},
  url = {http://dx.doi.org/10.1140/epje/s10189-023-00334-2},
  DOI = {10.1140/epje/s10189-023-00334-2},
  number = {9},
  journal = {The European Physical Journal E},
  publisher = {Springer Science and Business Media LLC},
  author = {Schlaich,  Alexander and Tyagi,  Sandeep and Kesselheim,  Stefan and Sega,  Marcello and Holm,  Christian},
  year = {2023},
  month = sep,
  pages = {80}
}

@article{Trizac2003,
author = {Trizac, E. and Bocquet, L. and Aubouy, M. and von Gr{\"u}nberg, H. H.},
title = {Alexander's Prescription for Colloidal Charge Renormalization},
journal = {Langmuir},
volume = {19},
number = {9},
pages = {4027-4033},
year = {2003},
doi = {10.1021/la027056m},
URL = {https://doi.org/10.1021/la027056m},
}

@article{Denton2000,
  title = {Effective interactions and volume energies in charged colloids: Linear response theory},
  volume = {62},
  ISSN = {1095-3787},
  url = {http://dx.doi.org/10.1103/PhysRevE.62.3855},
  DOI = {10.1103/physreve.62.3855},
  number = {3},
  journal = {Physical Review E},
  publisher = {American Physical Society (APS)},
  author = {Denton,  A. R.},
  year = {2000},
  month = sep,
  pages = {3855–3864}
}

@article{terRele2026,
  title = {Machine-learned many-body potentials for charged colloids reveal gas–liquid spinodal instabilities only in the strong-coupling regime of primitive models},
  volume = {164},
  ISSN = {1089-7690},
  url = {http://dx.doi.org/10.1063/5.0318479},
  DOI = {10.1063/5.0318479},
  number = {14},
  journal = {The Journal of Chemical Physics},
  publisher = {AIP Publishing},
  author = {ter Rele,  Thijs and van Roij,  René and Dijkstra,  Marjolein},
  year = {2026},
  month = apr, 
  pages = {144102}
}

@article{Gouy1910,
  title = {Sur la constitution de la charge électrique à la surface d’un électrolyte},
  volume = {9},
  ISSN = {0368-3893},
  url = {http://dx.doi.org/10.1051/jphystap:019100090045700},
  DOI = {10.1051/jphystap:019100090045700},
  number = {1},
  journal = {Journal de Physique Théorique et Appliquée},
  publisher = {EDP Sciences},
  author = {Gouy,  M.},
  year = {1910},
  pages = {457–468}
}

@article{Chapman1913,
  title = {LI. A contribution to the theory of electrocapillarity},
  volume = {25},
  ISSN = {1941-5990},
  url = {http://dx.doi.org/10.1080/14786440408634187},
  DOI = {10.1080/14786440408634187},
  number = {148},
  journal = {The London,  Edinburgh,  and Dublin Philosophical Magazine and Journal of Science},
  publisher = {Informa UK Limited},
  author = {Chapman,  David Leonard},
  year = {1913},
  month = apr,
  pages = {475–481}
}

@article{baumgartl-2006,
	author = {Baumgartl, Jörg and Arauz-Lara, Jose Luis and Bechinger, Clemens},
	journal = {Soft Matter},
	month = {1},
	number = {8},
	pages = {631},
	title = {{Like-charge attraction in confinement: myth or truth?}},
	volume = {2},
	year = {2006},
	doi = {10.1039/b603052a},
	url = {https://doi.org/10.1039/b603052a},
}

@article{filion-2011,
	author = {Filion, Laura and Ni, Ran and Frenkel, Daan and Dijkstra, Marjolein},
	journal = {The Journal of Chemical Physics},
	month = {4},
	number = {13},
    	pages = {134901},
	title = {{Simulation of nucleation in almost hard-sphere colloids: The discrepancy between experiment and simulation persists}},
	volume = {134},
	year = {2011},
	doi = {10.1063/1.3572059},
	url = {https://doi.org/10.1063/1.3572059},
}

@article{Trefalt2015,
  title = {Charge Regulation in the Electrical Double Layer: Ion Adsorption and Surface Interactions},
  volume = {32},
  ISSN = {1520-5827},
  url = {http://dx.doi.org/10.1021/acs.langmuir.5b03611},
  DOI = {10.1021/acs.langmuir.5b03611},
  number = {2},
  journal = {Langmuir},
  publisher = {American Chemical Society (ACS)},
  author = {Trefalt,  Gregor and Behrens,  Sven Holger and Borkovec,  Michal},
  year = {2015},
  month = Dec,
  pages = {380–400}
}

@misc{comsol,
    title={{COMSOL Multiphysics® User's Guide, Version 5.6}},
    year={2025},
    publisher={COMSOL AB}
}

\clearpage
\appendix
\section{Molecular Dynamics Simulations}

The simulations presented in this work, both in the primitive model  and those  using the ML potential, are molecular dynamics (MD) simulations. All simulations were performed using the LAMMPS software package~\cite{LAMMPS} at constant volume $V$ and temperature $T$. The equations of motion are integrated using a velocity-Verlet scheme, while a Nos\'{e}-Hoover thermostat is employed to maintain an average  kinetic energy per particle $\langle E_{kin} \rangle = 3 k_B T/2$. 

The simulations using  the ML potential are performed at constant number of colloids $N$, i.e. in the $NVT$-ensemble. The ML potentials were implemented in LAMMPS  using the High-Dimensional Neural Network Potentials (HDNNP) package.\cite{Singraber2019} For more information on ML potential simulations, see Ref. \citenum{terRele2025}. 

In  PM simulations, the number of colloids is kept fixed at $N$, while the ions are treated grand-canonically, corresponding to a semi-grand canonical ensemble. Coion-counterion pairs are inserted and removed via grand-canonical Monte Carlo (GCMC) moves with an associated chemical potential $\mu$. 
All particles--the colloids and the ions--have the same mass $m$. The colloid radius is 20 times larger than the ion radius, i.e. $a_i/a = 1/20$. From these parameters, we define  the characteristic MD timestep $\tau_{MD} = \sqrt{4m a^2/(k_B T)}$. The integration timestep used in all simulations is equal to $\Delta t =  0.0005 \tau_{MD}$. 
Long-range electrostatic interactions are computed using the particle-particle particle-mesh (PPPM) method with a target relative error of $10^{-4}$.\cite{Hockney1988} 

\section{Symmetry Functions and Gradients}

In  Section \ref{sec:ML-procedure}, we describe how colloidal forces are fitted to a ML potential using a procedure previously introduced in  Ref. \citenum{terRele2025}. The forces are expressed in terms of a set of two-body and three-body symmetry functions introduced by Behler and Parrinello.\cite{Behler2007} The two-body symmetry function is defined as \begin{align}\label{G2}
    G^{(2)}(i) &= \sum_j e^{-\gamma(R_{ij} - R_s)^2} f_c(R_{ij}). 
\end{align} Here, $R_{ij} = |\textbf{R}_i - \textbf{R}_j| $ denotes the radial distance between colloid $i$ and $j$, and the sum runs over all neighboring particles $j$.
The parameters $\gamma$ and $R_s$ are optimization parameters that control the width and center of the Gaussian in $R_{ij}$, respectively. The two-body symmetry function includes a cut-off function $f_c(R_{ij})$, defined as   
\begin{equation}
    f_c(R_{ij}) =  \begin{cases}
\tanh^3{\left(1 - R_{ij}/R_c\right)} & \text{for } R_{ij} \leq R_c; \\
0 &\text{for } R_{ij} > R_c,
\end{cases}
\end{equation}
which smoothly decays to zero as $R_{ij}$ approaches the  cut-off distance $R_c$. The three-body symmetry function is defined as
\begin{align}\label{G3}
    G^{(3)}(i) = & 2^{1-\xi} \sum_{j, k\neq i} \left(1 + \lambda \cos{\theta_{ijk}} \right)^{\xi} \nonumber
    \\ & e^{-\gamma \left(R_{ij}^2 + R_{ik}^2 + R_{jk}^2\right)} f_c(R_{ij}) f_c(R_{jk}) f_c(R_{ik}).
\end{align} 
For the two-body symmetry functions $G^{(2)}$, the optimization parameters are: 

 \noindent $\gamma a^{2} \in \{0.0025, 0.025, 0.25, 0.5, 1, 2, 4\}$;

\noindent$R_s/a \in \{0.0, 0.2, 0.4, 0.6, 0.8, 1.0, 1.2, 1.4, 1.6, 2.0, 4.0\}$.  

\noindent For the three-body symmetry functions $G^{(3)}$, the parameters are: 

\noindent $\gamma a^{2} \in \{0.0025, 0.025, 0.25, 0.5, 1, 2, 4\}$; 

\noindent $\lambda \in \{1, 2, 4, 8, 16, 32\}$; $\xi \in \{1, -1\}$. 

\noindent Throughout this work, we use a cut-off radius  $R_c = 8.0 a$. 
For additional details on the construction of the ML potentials and the  symmetry-function framework, we refer the reader to Ref. \citenum{terRele2025}.

\section{Bulk potential Z=1200 }\label{sec:app:bulkpotential}
\begin{figure}[h]
\centering
\includegraphics[width=0.99\linewidth]{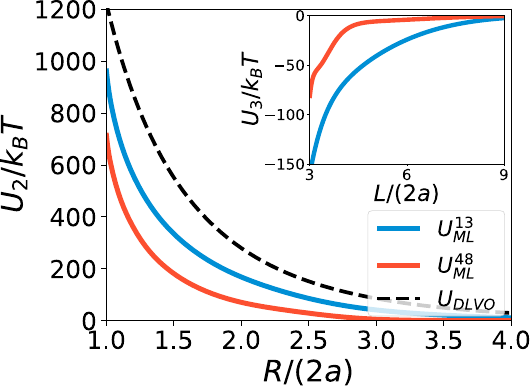}
\caption{\textbf{ML potential $\beta U^{48}_{ML}$ trained on Poisson-Boltzmann calculations of 48 charged colloids} at effective temperature $a/\lambda_B =92.7 $, screening parameter $\kappa a  = 0.392$, and charge valency $Z = 1200$. For training,  only the  inner 13 ``bulk'' colloids are selected. A comparison is made to the potential trained on 13 isolated colloids $\beta U^{13}_{ML}$.  The two-body contribution of the ML potential $\beta  U^{(2)}_{ML}$ is shown in the main panel, while the three-body contribution $ \beta U^{(3)}_{ML}$ is shown in the inset. For the DLVO potential, a renormalized charge valency $Z^* = 663$ is used. }
\label{fig:13_48_comparison_Z1200}
\end{figure}

We employ the technique described in Section \ref{sec:48colloids} to generate an ML potential  $\beta U_{ML}^{48}$  for colloids with constant charge valency $Z=1200$ at effective temperature $a/\lambda_B = 92.7$, and screening parameter $\kappa a = 0.392$, again using the forces acting on the inner 13 colloids of 48-colloid clusters as training input.  In Figs. \ref{fig:twocol_pot} and \ref{fig:threecol_pot_shape}, we have already presented the grand potential contributions $\beta \Omega^{(2)}$ and $\beta \Omega^{(3)}$ for this parameter set, obtained from finite-element Poisson-Boltzmann calculations of isolated colloid pairs and triplets, respectively. In Fig. \ref{fig:13_48_comparison_Z1200}, we present the pairwise (main figure) and three-body contributions (inset) to $U_{ML}^{48}$ (orange) and $U_{ML}^{13}$ (blue) for this system. For reference, the pairwise DLVO potential (dashed line) is also shown.  We observe, in agreement with the results for $Z=100$ in the main text, that $U_{ML}^{48}$ is weaker than $U_{ML}^{13}$, with the latter itself being  weaker than the DLVO potential for pair interactions. Thus the ``best'' inclusion of bulk-like many-body conditions, $U_{ML}^{48}$, leads to the weakest effective many-body contributions, in particular the weakest contribution to the cohesive energy that is needed to stabilize coexistence between a dense colloidal liquid or crystal with a dilute colloidal gas phase. 
Indeed, Molecular Dynamics simulations  performed with $U_{ML}^{48}$ do not exhibit colloidal phase separation. Hence, the results for colloids with  valency $Z = 1200$ are therefore consistent with those obtained for  colloids with valency $Z =100$, as discussed in Section \ref{sec:48colloids}. 

\section{DLVO potential with 3-body correction}

\begin{figure*}[h]
\centering
\includegraphics[width=0.8\linewidth]{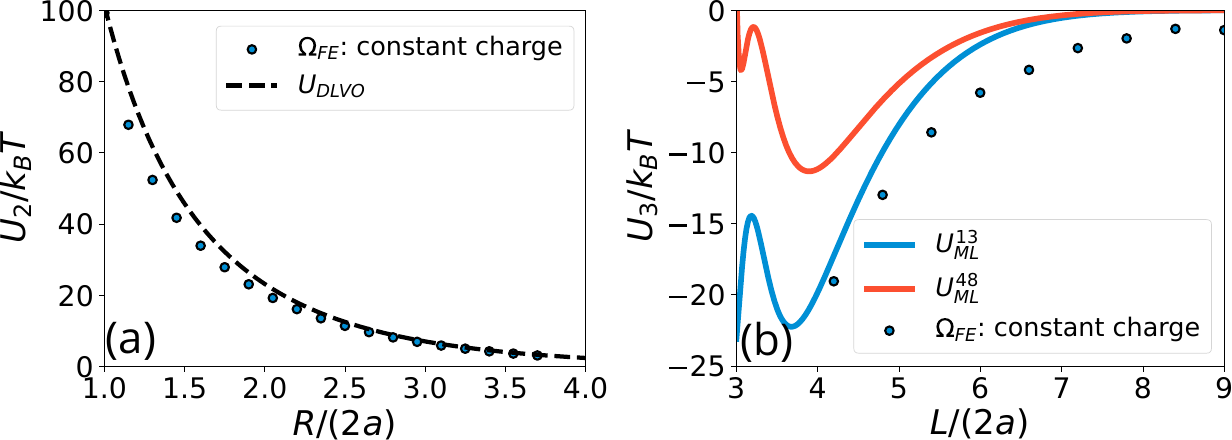}
\caption{\textbf{Three-body ML correction $\beta U^{48}_{ML}$ to the DLVO potential} for charged colloids at effective temperature $a/\lambda_B =7.7 $, screening parameter $\kappa a  = 0.392$, and charge valency $Z = 100$. The potential correction $\beta U^{48}_{ML}$ is trained on Poisson–Boltzmann calculations of 48-colloid clusters, using only the inner 13 ``bulk'' colloids for training.  (a) The two-body DLVO potential $U^{}_{DLVO}$, calculated using the renormalized charge valency $Z^* = 55$, together with the grand potential $\Omega^{(2)}_{FE}$ for an isolated pair of colloids.  (b) The three-body potential correction $ \beta U^{(3)}_{ML}$, obtained from  the machine-learning procedure, together with a comparison to the potential $\beta U^{13}_{ML}$ trained on isolated clusters of 13 colloids. }
\label{fig:3-body-fit}
\end{figure*}

In the preceding sections, we fitted the forces acting on the colloids using a combination of two- and three-body symmetry functions. However, the interaction potential between isolated colloid pairs is already described very accurately by the DLVO potential, as was demonstrated in Figs. \ref{fig:twocol_pot}(b) and \ref{fig:ML_PB_comparison}(a). To more directly isolate the many-body components of the interaction potential, we fix the two-body interaction to the DLVO potential, given in Eq. (\ref{eq:DLVO}), and use the ML procedure to fit the remaining many-body contribution using a set of three-body symmetry functions. In Fig. \ref{fig:3-body-fit} we present the resulting potential for colloids of charge valency $Z = 100$, effective temperature $a/\lambda_B = 7.7$, and screening parameter $\kappa a  = 0.392$. Similar to Section \ref{sec:48colloids}, we compare a potential trained on PB calculations of 13 isolated colloids to a potential trained on the inner 13 colloids of a 48-colloid cluster. However, in this section the two-body interaction is fixed to the DLVO potential, which we compare to the two-body grand potential $\Omega^{(2)}_{FE}$ in Fig. \ref{fig:3-body-fit}(a). The three-body correction to this DLVO potential is plotted in Fig. \ref{fig:3-body-fit}(b). As in Fig. \ref{fig:13_48_comparison}, we  find that the cohesive contribution of the many-body correction is weaker when the potential is trained on forces from colloids in bulk-like environments  ($\beta U^{48}_{ML}$) compared with training on finite 13-colloid clusters ($\beta U^{13}_{ML}$). These results are consistent with those presented in Section \ref{sec:48colloids}, indicating that the many-body correction to the  DLVO pair potential is attractive. However, we again find that this attraction weakens as the training environment more closely approaches bulk conditions.

\end{document}